\documentclass{article}

\usepackage{arxiv}
\usepackage{authblk}
\usepackage[utf8]{inputenc} % allow utf-8 input
\usepackage[T1]{fontenc}    % use 8-bit T1 fonts
\usepackage{hyperref}       % hyperlinks
\usepackage{url}            % simple URL typesetting
\usepackage{booktabs}       % professional-quality tables
\usepackage{amsfonts}       % blackboard math symbols
\usepackage{nicefrac}       % compact symbols for 1/2, etc.
\usepackage{microtype}      % microtypography
\usepackage{cleveref}       % smart cross-referencing
\usepackage{lipsum}         % Can be removed after putting your text content
\usepackage{graphicx}
\usepackage{doi}
\usepackage{multirow}
\usepackage{hyperref}
\usepackage{longtable}
\usepackage{array}
\usepackage[utf8]{inputenc}
\usepackage{booktabs}
\usepackage[natbib=true, backend=biber, bibstyle=numeric, sorting=none]{biblatex}

\newcolumntype{P}[1]{>{\raggedright\arraybackslash}p{#1}}
\newcolumntype{C}[1]{>{\centering\arraybackslash}p{#1}}

\title{Focus Areas, Themes, and Objectives of Non-Functional Requirements in DevOps: A Systematic Mapping Study}

\date{}

\AtEveryBibitem{%
    \ifentrytype{inproceedings}{%
    	\clearfield{url}%
    	\clearfield{urldate}%
    	\clearfield{month}%
    	\clearfield{notes}%
    	\clearfield{issn}%
    }{}%
    \ifentrytype{inproceedings}{%
    	\clearfield{url}%
    	\clearfield{urldate}%
    	\clearfield{month}%
    	\clearfield{notes}%
    	\clearfield{issn}%
    }{}%
}

\DeclareSourcemap{
  \maps[datatype=bibtex]{
    \map{
      \step[fieldset=issn, null]
      \step[fieldset=month, null]
      \step[fieldset=notes, null]
      \step[fieldset=urldate, null]
    }
  }
}

\author[1,2]{Philipp Haindl}
\author[2]{Reinhold Plösch}
\affil[1]{Software Competence Center Hagenberg, Austria}
\affil[2]{Department of Business Informatics - Software Engineering, Johannes Kepler University Linz, Austria}

\renewcommand{\headeright}{}
\renewcommand{\undertitle}{Technical Report}
\renewcommand{\shorttitle}{Non-Functional Requirements in DevOps: A Systematic Mapping Study}

\hypersetup{
pdftitle={A template for the arxiv style},
pdfsubject={q-bio.NC, q-bio.QM},
pdfauthor={Philipp Haindl, Reinhold Plösch}
}

\bibliography{manuscript.bib}  
\begin{document}

\maketitle

\begin{abstract}
\textbf{Context:} Software non-functional requirements address a multitude of objectives, expectations, and even liabilities that must be considered during development and operation. Typically, these non-functional requirements originate from different domains and their concrete scope, notion, and demarcation to functional requirements is often ambiguous. \textbf{Objective:} In this study we seek to categorize and analyze relevant work related to software engineering in a DevOps context in order to clarify the different focus areas, themes, and objectives underlying non-functional requirements and also to identify future research directions in this field. \textbf{Method:} We conducted a systematic mapping study, including 142 selected primary studies, extracted the focus areas, and synthesized the themes and objectives of the described NFRs. In order to examine non-engineering-focused studies related to non-functional requirements in DevOps, we conducted a backward snowballing step and additionally included 17 primary studies. \textbf{Results:} Our analysis revealed 7 recurrent focus areas and 41 themes that characterize NFRs in DevOps, along with typical objectives for these themes. Overall, the focus areas and themes of NFRs in DevOps are very diverse and reflect the different perspectives required to align software engineering with technical quality, business, compliance, and organizational considerations. \textbf{Conclusion:} The lack of methodological support for specifying, measuring, and evaluating fulfillment of these NFRs in DevOps-driven projects offers ample opportunities for future research in this field. Particularly, there is a need for empirically validated approaches for operationalizing non-engineering-focused objectives of software.
\end{abstract}

% keywords can be removed
\keywords{Non-functional Requirements \and Software Quality Attributes \and DevOps}

\section*{Remark}
This paper is an addition to our peer-reviewed publication in \cite{9226285} and contains a more elaborate presentation of our findings. When citing our work, please always cite the peer-reviewed publication; citing this technical report should always be optional for cases where you explicitly reference aspects that were not published in the peer-reviewed publication.

\section{Introduction}
\label{sec:introduction}
The common notion of non-functional requirements (NFRs) of software,
such as e.g., maintainability or performance efficiency, primarily
targets \emph{engineering-focused} objectives of software. As such they
disregard the also \emph{non-engineering-focused} objectives pursued with
software, e.g., that it is frequently used by customers, generates
revenue aligned with the business model, and that its monitored usage
data allows to ideate new features for the customers. Due to the lack of
awareness for these non-engineering-focused NFRs, they are often only
vaguely elicited and balanced with functional requirements \cite{chung_non-functional_2000,chung_non-functional_2009,daneva_software_2013,doerr_non-functional_2005}. This
in turn disguises business, staffing, and organizational considerations
that also have an impact on decisions in software projects. Also, it
makes it difficult for software engineers to appropriately react to
deviations of these NFRs in DevOps. Due to the absence of a
comprehensive classification scheme that takes into account the
heterogeneity of domains having an impact on NFRs, it is cumbersome to
properly approach them in software specification documents and meet the
different stakeholders' objectives. As an example, the objectives
towards exchangeability of software developers or transaction-cost
monitoring in DevOps exceed the primarily engineering-focused
understanding of NFRs. These cases of NFRs are primarily applicable in
DevOps as they require a thorough indentation between the development
and operation perspective of software.

Specifically the evolution of BizDevOps \cite{gruhn_bizdevops_2015,forbrig_integrating_2019} emphasizes the necessity for authors to more intensively integrate business and other non-engineering-focused NFRs in
the DevOps context. Several works have stressed the required alignment
between business processes, strategic objectives, company culture, and
engineering decisions with DevOps \cite{jabbari_what_2016,senapathi_devops_2018,beulen_implementing_2018,hemon_agile_2019}. In addition, the ubiquity and volume of data accruing in DevOps provide ample methods for evaluating
fulfillment of NFRs to continuously improve the software. The absence of
a classification scheme for NFRs in DevOps and the thereof pursued
objectives also hinder the effective formulation of respective measures
that can then be acquired and assessed in the DevOps cycle.

\subsection{Problem statement and motivation}
\label{sec:motivation}
To increase the value of DevOps in practice and assure a better
alignment with non-engineering-focused objectives in software projects, a comprehensive classification of NFRs, typical themes, pursued objectives and implementation practices in DevOps is required. In order to examine the handling of NFRs in DevOps we conducted a systematic mapping study in combination with backward snowballing that included 142 primary studies. The following paper presents the findings from this mapping study.\

The focus of this research was on examining
\begin{itemize}
    \item which NFRs are relevant and typically handled in DevOps, and
    \item how these NFRs can be classified by focus areas and subordinate themes, and
    \item what objectives are pursued with these NFRs;
\end{itemize}
We also analyzed the publication channels, contributions and research
facets as well as the frequency distribution over the years. In order to
get a solid understanding of the current state of practice of NFRs in
DevOps, we followed a snowballing procedure and selected further
relevant papers for inclusion in our analysis. In the context of this
paper we refer to DevOps as \textit{"a set of practices intended to reduce the
time between committing a change to a system and the change being placed
into normal production while ensuring quality"} \cite{bass_devops_2015}. The
introduction and practice of DevOps shall however not be seen as an
isolated software engineering activity, but also requires organizational
changes. In many cases this organizational change is much more
fundamental than the changes in the software engineering process, as it
additionally requires significant adaption to other processes, e.g.,
change management processes \cite{tamburri_architects_2016}.

Our work seeks to augment the common understanding of NFRs by
contributing new facets about the different scopes and objectives behind
NFRs, specially in DevOps. The different perception and understanding of
NFRs also becomes apparent when skimming the definitions of NFRs from
other publications. In one of the first publications in this domain
\cite{freeman_software_1987}, economic aspects of software are mentioned in the same
turn as purely engineering-related aspects such as reliability,
usability or safety. Other authors \cite{liu_integrating_2010,broy_rethinking_2015,eckhardt_are_2016} reflect only on the structural and behavioral aspects of NFRs and hence also approach them solely from a technical perspective.
In contrast, recent works in this context \cite{ameller_dealing_2010,hiisila_combining_2015} have a broader notion of NFRs. These works stress that
also \textit{"...environmental factors, such as legislation related to the
business branch and compliance, may introduce vital requirements..."} and
that \textit{"...user needs and other stakeholders should not be forgotten,
either."}. To summarize, there still is no clear demarcation between
functional and non-functional requirements \cite{glinz_non-functional_2007,broy_rethinking_2015} and
how to approach and evaluate non-engineering-focused NFRs that affect
the design, implementation, and operation of software.

\subsection{Contribution}\label{contribution}
Our study shows that there are multiple perspectives on \textit{why} and \textit{how} 
to approach NFRs in DevOps. These perspectives are reflected by
different points of focus that are typically (and often tacitly) taken
in software projects and range from engineering-related qualities, e.g.,
static or dynamic code analysis, to non-engineering-related qualities
addressing business, legal, or organizational requirements. As an
example, these requirements address how to balance economic requirements
of the software, the mapping of software artefacts and their
contribution to revenue streams, or motivational factors influencing the
development and operation of software.

Resulting from our analysis of 142 relevant papers in this systematic
mapping study, we contribute an analysis and thematic classification of
the current state of research in this field. The analysis of the studies
takes into account the research and contribution facets, publication
channels, and the publication frequency of the studies over time.
Complementary, the NFRs in DevOps described in the studies are
classified by focus area, typical themes, and pursued objectives. In
particular, our work seeks to extend the primarily engineering-focused
understanding of NFRs by showing their heterogeneity in the analyzed
studies and their relevance specially in DevOps.

\subsection{Outline}\label{sec:outline}
The remainder of the paper is organized as follows: In Section \ref{sec:related_work} we
provide an overview of work related to our study. Section \ref{sec:research_methodology} presents the research methodology, introducing the research goals and questions and
how each method has been applied in this literature study. The analysis
of the study population showing the different facets of the studies,
publication channels, and the publication frequency over time is
elaborated in Section \ref{sec:overview_rq1}. Following, the focus areas of NFRs in DevOps are presented in Section \ref{sec:focus_areas_rq2}, before the themes and objectives related to these NFR focus areas in DevOps are portrayed in Section \ref{sec:themes_rq3}. The findings from our study are discussed in Section \ref{sec:discussion}, followed by outlining the threats to the validity of our results in Section \ref{sec:threats}. Finally, we conclude our paper in Section \ref{sec:conclusion} and attach the details of the used classification schema in \ref{appendix:appendix_a} and the systematic map of all analyzed primary studies in \ref{appendix:appendix_b}.
\section{Related work}
\label{sec:related_work}

In this section we elaborate on related work in the context of handling
NFRs in DevOps, the necessity for this study and how we demarcate our
work from other authors. Basically, 3 streams of research can be
identified:

The first stream of related research deals with the scope and notion of
NFRs towards software. Particularly, this tackles the question where to
draw the demarcation between non-functional and functional requirements
and whether the notion of NFRs also covers non-engineering-related
concerns, e.g., business, organizational or legal concerns. These
concerns also affect the design, development, and operation of the
software \cite{glinz_non-functional_2007,chung_non-functional_2009,broy_rethinking_2015} and may also
become manifest in NFRs. The differentiation between non-functional and
functional requirements is not only prevalent in research, but also
affects the elicitation, documentation, and validation of NFRs in
software projects. Eckhardt et al. \cite{eckhardt_are_2016} analyzed 530 NFRs of 11
industrial requirements specifications regarding the extent these NFRs
describe system behavior. They concluded that most many so-called NFRs
actually are \textit{not} non-functional, but instead could be treated as
functional requirements. However, the focus of the analyzed NFRs is
primarily engineering-oriented and leaves out non-engineering-related
concerns which may also become manifest in NFRs. Mairiza et al. \cite{mairiza_investigation_2010}
examined 182 sources of information elaborating NFRs, ranging from
academic articles to technical reports and white papers, and categorized
them depending on their type of being either business, external,
development or quality related. For each extracted NFR from these
sources, they analyzed whether it adheres to a common definition, has
concrete measurable attributes and a concrete measurement methodology
defined for it. They concluded that only around 20\% of NFRs typically
arising in software projects are commonly defined covering attributes
and measurement methodology. The remaining 26\% of NFRs at least have a
definition, but for the majority of over 53\% of common NFRs neither
their attributes nor their measurement methodology is defined, leaving
them factually ignored in the development process. Also, the authors of
this work conclude that the common notion of NFRs leaves out important
non-engineering-related aspects that have an impact specially on the
development of the software and specifically, that further research
addressing scope, notion, and measurement of NFRs shall be done.

The second stream of relevant works comprises systematic literature
reviews and mapping studies. Hasan et al. \cite{hasan_classification_2014} argue that NFRs are often ignored, inadequately specified, and rarely treated as first-class
elements such as functional requirements. In the frame of a systematic
literature review (SLR) the authors analyzed 92 describing approaches
for handling NFRs in the software development lifecycle, to examine the
documentation approaches and scopes of NFRs in software projects. The
focus of their work is not on the scope of NFRs itself, but primarily on
approaches for their elicitation, documentation, and handling. The
authors categorize these approaches into goal-, aspect-, or
pattern-oriented NFR elicitation approaches. While goal-oriented NFR
elicitation techniques result in higher preciseness and less ambiguity
of the NFR specification, the authors also stress that the plethora of
NFR elicitation techniques is due to the lack of a thorough taxonomy for
NFRs. They also underline that despite the undisputed advantages of
these techniques, most of them can only be used by domain experts, which
complicates NFR elicitation and specification in software projects.
Ouhbi et al. \cite{ouhbi_software_2013} performed a systematic mapping study based on 51
primary studies, mainly to analyze the population of studies dealing
with software quality elicitation and specification. In their study, the
authors solely take into account NFRs that adhere to the ISO/IEC 25010
specification. Their study summarizes that NFRs are mainly validated
during development and that data from software operation are often
disregarded or not used effectively for NFR validation. Also, the
authors conclude that there is a need for more studies focusing on the
validation of NFRs throughout the development lifecycle.

The last stream of relevant research examines the handling of NFRs in
the DevOps context. In a qualitative multiple-case study,
Riungu-Kalliosaari et al. \cite{riungu-kalliosaari_devops_2016} interviewed practitioners about the challenges and benefits of DevOps adoption. In the interviews, the
respondents mentioned that NFRs in DevOps primarily focus on product
quality, thus neglecting the quality of the software development,
delivery, and operation process. This aligns with the work of Senapathi
et al. \cite{senapathi_devops_2018} who conducted an exploratory case study interviewing
several software engineers to explore the challenges linked with
adopting DevOps practices. In the interviews the experts responded that
in practice they are confronted with non-engineering-focused NFRs such
as e.g., the alignment of the software with revenue streams or feature
ideation from monitored user behavior. These non-engineering-focused
NFRs are not fully covered by the primarily engineering-oriented notion
of NFRs and hence are often neglected in DevOps. Though, similar to
engineering-focused NFRs such as e.g., maintainability of the source
code, they have a direct effect on whether the objectives pursued with
software can be fulfilled and with what efforts and costs.\\

It can be summarized that multiple related works stress the solely
technical notion of NFRs and suggest to rethink their scope to more
effectively specify, measure, and evaluate non-engineering-focused NFRs,
e.g., business, organizational, human, and legal requirements.
Particularly in DevOps with a multitude of data originating from
different systems, a thorough catalogue of NFRs along with their focus,
typical themes, subordinate objectives, and measures is the prerequisite
for more effectively approaching NFRs. The presented systematic mapping
study aims to close this gap through providing a detailed classification
of NFRs in DevOps -- by focus areas, themes, and pursued objectives.
\section{Research methodology}
\label{sec:research_methodology}

The research presented in this paper resulted from a systematic mapping
study conducted over the period of 9 months (December 2018 to August
2019). There is an important distinction between systematic literature
reviews and mapping studies: Systematic literature reviews (SLRs) intend
to \textit{"identify best practice with respect to specific procedures,
technologies, methods or tools by aggregating information from
comparative studies"} while the result of a systematic mapping studies
is a \textit{"classification and thematic analysis literature on a software
engineering topic"} \cite{kitchenham_using_2011}. As already outlined in
Section \ref{sec:related_work}, the notion and scope of NFRs still remain vague in academic
works and industry, specifically whether non-engineering-focused
concerns are also covered by this term. Therefore, in our mapping study
we augmented the set of analyzed papers through a snowballing procedure
\cite{wohlin_guidelines_2014} to also capture primary studies that elaborate on
non-engineering-focused NFRs in DevOps. For the sake of transparency and
reproducibility of our study, we aimed to keep the number of additionally included
papers small.

\subsection{Research goal and questions}
\label{sec:research_questions}

The overall research goal for this study can be described through the
following objectives: to (a) analyze the publication population in this
field and particularly the research and contribution facets; (b) provide
a comprehensive catalogue of engineering- as well as 
non-engineering-focused NFRs; (c) nurture a broader understanding of
the different focus areas, themes, and subordinate objectives of NFRs in
DevOps; and (d) outline directions for future research. Below we present
the research questions that were derived from these objectives.
\begin{itemize}
\item   \textit{What is the current state of research pertaining to NFRs in the context of DevOps? (RQ1)}

We took common bibliometric perspectives onto the current state of
research pertaining to this field and refined them into corresponding
research questions.
\begin{itemize}
	\item What is the distribution of research facets in studies related to NFRs in DevOps? (RQ1.1)
	\item What is the distribution of contribution facets in these studies? (RQ1.2)
	\item What are the publication channels and venues used to publish related studies? (RQ1.3)
	\item How did the publication frequency of related studies evolve over time? (RQ1.4)
\end{itemize}
In RQ1.1 and RQ1.2 the perspectives are on the research and
contribution facets of the works respectively. Complementary, RQ1.3
examines the typical publication channels and venues of works in the
field of NFRs in DevOps. The analysis of the respective publication
population is finally supplemented by RQ1.4, which investigates how the
publication frequency of the studies evolved over time.
\item   \textit{How can NFRs typically relevant in the context of DevOps be
    characterized? (RQ2)} 
    
The focus of RQ2 is on carving out which NFRs can be discovered
recurrently among the different studies and are typically important in
DevOps. Therefore, the NFRs described in the studies are categorized by
their conjoint focus area, i.e., their spanning characteristics and
objectives.

\item   \textit{Which themes and objectives are associated with each focus area of
    NFRs in DevOps? (RQ3)}
    
This question analyzes representative themes and objectives that are
pursued with each NFR focus area. If possible, we present them along
with tangible practical examples and associated challenges.

\end{itemize}
Based on these research questions we defined the data to be extracted
from the primary studies (cf. Section \ref{sec:data_extraction}).

\subsection{Research process and steps}
\label{sec:research_process}

\begin{figure*}[t]
\centering
\includegraphics[scale=0.6]{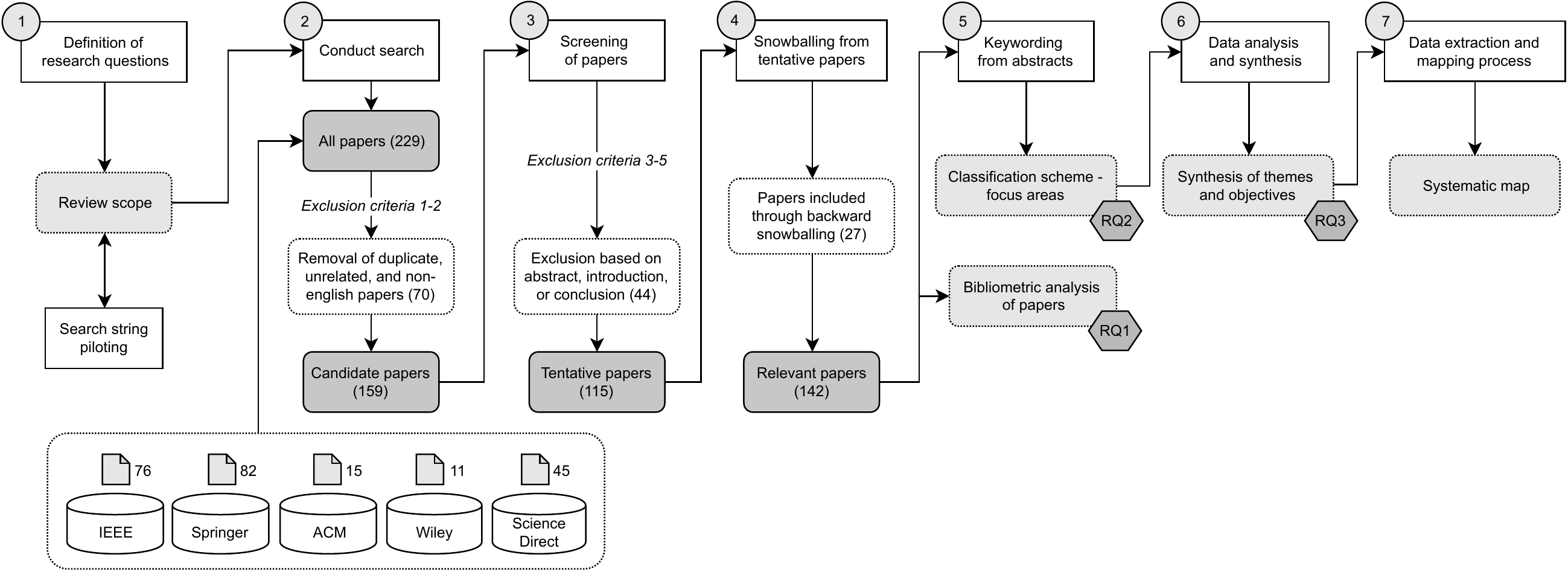}
\caption{Research process and steps of the mapping study, modified from Petersen et al. \cite{petersen_systematic_2008}.}
\label{fig:research_process}
\end{figure*}

The overall research design of our mapping study is depicted in Figure \ref{fig:research_process}. 
We followed the process as suggested by Petersen et al. \cite{petersen_systematic_2008} but
extended it by a snowballing step (step 4, Figure \ref{fig:research_process}). The process
itself was iterative and we repeated individual steps the more we
learned about the data available in the primary studies and how we could
integrate them in our study. Also, we refined the research questions
with increasing knowledge about NFRs, DevOps, and the challenges that
are associated with the combination of these two topics.

Below we describe the design and execution of the systematic mapping
study in a sequential manner: The need for the study emerged from our
related activities in this field \cite{haindl_towards_2019,haindl_extension_2019} 
pertaining to the extension of an operational software quality
model to evaluate the fulfillment of NFRs on the level of software
features, particularly in DevOps. The research questions (step 1,
Figure \ref{fig:research_process}) are elaborated in Section \ref{sec:research_questions}. They define the concrete
review scope of the systematic mapping study. Accordingly, the review
scope also influences the search string used to retrieve representative
papers for the topic of interest. We piloted the search string several
times until we ended up with a clear definition of it, which repeatedly
returned the most stable and complete set of relevant papers for this
topic.

Consequently, we searched for publications on selected databases (cf.
Section \ref{sec:searchstrategy}) using predefined search strings (step 2, Figure \ref{fig:research_process})
following the PICO schema \cite{kitchenham_guidelines_2007}. The concrete
search strategy, the database-dependent search strings, and the
rationales behind the composites of the search strings are described in Section \ref{sec:searchstrategy}. 
From all papers we excluded duplicate, unrelated, and
non-english works upfront (exclusion criteria 1-2). Then we screened
these \textit{candidate papers} thoroughly (step 3, Figure \ref{fig:research_process}) and again
selected only papers according to the predefined inclusion and exclusion
criteria \textit{(tentative papers)}. The applied inclusion and exclusion
criteria are detailed in Section \ref{sec:primary_study_selection}.

Several authors emphasize that the scope and notion of NFRs are still
vague in academic literature \cite{glinz_non-functional_2007,chung_non-functional_2009,broy_rethinking_2015}. 
To address this limitation and to capture further papers
contributing to a broader notion of NFRs, we subjected the tentative
papers (step 3, Figure \ref{fig:research_process}) a backward-snowballing procedure and
included selected cited papers (step 4, Figure \ref{fig:research_process}) into the final
paper set (relevant papers) for this mapping study. The snowballing
procedure \cite{wohlin_guidelines_2014} is elaborated in Section \ref{sec:snowballing}. Consequently,
we extracted recurrent keywords and key themes from the abstracts of the
papers to develop and repeatedly refined our classification schema (step
5, Figure \ref{fig:research_process}). Resulting from this step is a classification scheme that
particularly regards the focus areas of NFRs in DevOps, which is the
prerequisite for the systematic map as one outcome of the mapping study.
Then we analyzed all papers to gain a thorough understanding about these
NFRs (step 6, Figure \ref{fig:research_process}) to synthesize the focus areas, themes, and
objectives associated with NFRs in DevOps. In the last step we used the
classification scheme to extract the metadata, map the papers according
to predefined characteristics, e.g., research and contribution facets,
(step 7, Figure \ref{fig:research_process}). Eventually we built the systematic map to
illustrate the characteristics of the analyzed papers, the extracted
focus areas, themes, and the objectives of NFRs in DevOps.

\subsection{Search strategy and databases}
\label{sec:searchstrategy}
We conducted several experimental searches from December 2018 to January
2019 to gain a better understanding about the scope of research
available on the topic and the terms that are most frequently in the
respective works. For the experimental searches we used the pilot search
string \textit{"*DevOps" and ("non-functional" OR "quality")*"} on title,
keywords, and abstract and limited the searches to the IEEE database.
Based on this search string, we retrieved 69 papers of which we excluded
24 papers just by reading title or abstract. Subsequently, we read the
remaining 45 papers in detail and discovered that some of them reflected
on quality and NFRs in DevOps derivatives, e.g., BizDevOps or
Data-Driven DevOps. Thus, to capture a broader notion of the term
"DevOps" in these works, we adjusted the search strings accordingly
through wildcards. When developing the search string, we followed the
suggestions by Kitchenham and Charters \cite{kitchenham_guidelines_2007} and composed the search
string of several parts, representing either the population (of the
studied phenomenon) or the intervention to it. Table \ref{tab:pico} shows the parts
of the final search string along with their rationale for including
them.

\begin{table}[]
\centering
\begin{tabular}{p{80pt}p{120pt}p{230pt}}
\toprule
 & \textbf{Search term} & \textbf{Rationale} \\ \midrule
Population & *DevOps & Studies discussing DevOps principles and practices. The wildcard operator is used to also match related terms, e.g., in the context of "BizDevOps". \\
\hline
Intervention & qualit* & Studies discussing quality in the context of DevOps (e.g., "quality", "qualities"). \\
 & \multicolumn{2}{l}{OR} \\
 & non-functional* & Studies discussing non-functional requirements and similar terms in the context of DevOps (e.g., "non-functional attributes", "non-functional requirements", "non-functional concerns"). \\
 & \multicolumn{2}{l}{OR} \\
 & NFR* & Studies discussing non-functional requirements using the abbreviation "NFR".
\\
\bottomrule
\end{tabular}
\caption{Population, intervention, and search string keywords.}
\label{tab:pico}
\end{table}

During the pilot study we also discovered that though most works share
the common technical notion of NFRs, an ample amount of works elaborates
on non-engineering-focused NFRs. These works intensively contribute to a
broader understanding of NFRs and their objectives in general, which was
also described in Section \ref{sec:motivation} as research motivation for conducting this study. When analyzing the references of these papers, we also noticed
that they often referred to other relevant papers contributing to a
better understanding of the different types of NFRs. Thus, we decided to
subject these primary studies to a snowballing procedure as described in
Section \ref{sec:snowballing} and include their backward references if they meet the
defined inclusion requirements \cite{wohlin_guidelines_2014}.

The databases for searching the primary studies were selected
considering their coverage of the software engineering literature and
are shown along with the respective search string, filter, and retrieved
paper count in Table \ref{tab:databases}. In order to regard the individual syntax of each database and assure we retrieve the complete set of relevant works from
each database, we adapted the search string accordingly.\\

\begin{table}[]
\centering
\begin{tabular}{P{80pt}P{150pt}P{150pt}C{40pt}}
\toprule
\textbf{Database} & \textbf{Search string} & \textbf{Filter} & \textbf{Papers} \\ \hline
IEEE & ("*DevOps") AND ("qualit*" OR "non-functional*" OR "NFR*") & Only conference papers and journal articles & 76 \\ \hline
Springer & *DevOps ("qualit*" OR "non-functional*" OR "NFR*") & Only conference papers and journal articles in English in the following subject areas:  computer science, software engineering, and information systems applications & 82 \\ \hline
ACM Digital Library & +("*DevOps") +("qualit*" "non-functional*" "NFR*")) & None & 15 \\ \hline
Wiley & ("*DevOps") AND ("qualit*" OR "non-functional*" OR "NFR*") & None & 11 \\ \hline
Science Direct & DevOps AND ("qualit" OR "non-functional*" OR "NFR") & None & 45 \\ \hline
\textbf{Total} & & & \textbf{229} \\
\bottomrule
\end{tabular}
\caption{Selected databases and retrieved papers.}
\label{tab:databases}
\end{table}

Whenever possible we also defined filters to narrow the result set
according to the primary study selection criteria described in Section
\ref{sec:primary_study_selection}. We did not utilize an additional time filter as we were interested
in the full time spectrum of the studies. The top-right column of the
table shows the number of retrieved papers from each database, totaling
to 229 papers (up to and including July 2019).

\subsection{Primary study selection criteria}
\label{sec:primary_study_selection}
In Figure \ref{fig:research_process}, we illustrate the screening procedure, the numbers of
retrieved papers from each database, and the numbers of included and
excluded papers during study selection. The selection of the primary
studies was performed by the two co-authors with one researcher checking
the retrieved 229 primary studies prior to the screening (step 2, Figure \ref{fig:research_process}). 
This initial review comprised only title, author, and publication
channel of the papers. Inappropriate or duplicate papers were excluded
upfront, resulting in 70 of the 229 studies being excluded. To mitigate
study selection bias, we discussed disagreements of upfront excluded
papers until consensus was reached. Consequently, the remaining 159
primary studies were reviewed by both researchers against predefined
inclusion/exclusion criteria (step 3, Figure \ref{fig:research_process}). 
Both researchers reviewed all studies individually, resulting in 23 disagreements
regarding the selection of these studies. Again, we discussed the
selection of these studies in detail until we agreed on a final set for
the selection. Below we outline the applied inclusion and exclusion
criteria.\\

\noindent \textit{Inclusion criteria:}
Works such as books, book chapters, peer-reviewed studies published in
journals and conferences, and workshop papers reflecting, investigating,
or reporting on
\begin{enumerate}
    \item Elicitation, specification, operationalization, and monitoring of NFRs in DevOps
    \item Fulfillment evaluation of NFRs or quality aspects in DevOps
    \item Different engineering strategies and their effect on software quality and NFRs in DevOps
    \item The integration of technical and business objectives and their handling in DevOps
    \item Benefits, challenges, and limitations of DevOps and their effect on software quality
\end{enumerate}

\noindent \textit{Exclusion criteria:}
\begin{enumerate}
    \item Duplicate articles
    \item Works not written in English
    \item PhD or Master Theses
    \item Technical reports or white papers
    \item Works that not clearly discuss DevOps and NFRs in the frame of software engineering
\end{enumerate}

Though they are not peer-reviewed works, we also included books and book chapters retrieved from the academic databases in the initial screening. However, we only regarded these works if they could be attributed to the common body of knowledge or have at least one author with a strong and traceable academic record. In total we excluded 44 primary studies, the majority of the excluded papers being too generic or non-relevant to the studied topic. Our review resulted in 115 primary studies being selected for the subsequent snowballing procedure. Finally, we included additional 27 publications (of which 13 are books and book chapters) referenced by the 115 primary studies, which totals to 142 publications finally selected for the mapping study.

\subsection{Snowballing procedure}
\label{sec:snowballing}

As we discovered that also the references retrieved through the
systematic search contain valuable information for understanding the
notion and scope of NFRs, we decided to also include a backward
snowballing procedure. For the sake of transparency and repeatability of
our study, we paid attention on the number of papers added through
snowballing. Overall, we strived to keeping the number of thereby added
papers as little as possible, but also as extensive as necessary to
contribute to a broader notion specially of non-engineering-focused NFRs
in DevOps.

The snowballing procedure was done by two researchers conjointly, with
each one reviewing each paper separately and deciding about the further
handling of each paper again conjointly. If there were differences about
how to proceed with a paper, we repeated the analysis of the respective
paper and discussed until consensus was reached.

We followed the process as outlined by Wohlin \cite{wohlin_guidelines_2014} and selected the
115 papers remaining after exclusion based on abstract, introduction, or
conclusion (cf. Figure \ref{fig:research_process}) as the tentative start set for the snowballing
procedure. Thus, this set originated from the structured search on 5
academic databases and the inclusion and exclusion of papers adhered to
concrete and predefined criteria described in the previous section. In
total, we conducted 4 iterations of the snowballing process and thereby
analyzed each paper in its entirety and not only selected parts of it.
The reason for that being to evaluate if the NFRs described in an
analyzed paper do require additional complementary information, such as
most likely in the case of non-engineering-focused NFRs, or if the
described NFRs rather fit to the common and technical notion of NFRs,
thus not requiring any further elaboration.

As suggested by Wohlin \cite{wohlin_guidelines_2014}, when deciding whether to include a
referenced paper, we specially regarded the place and context in which
it was cited in the original paper, as this might give additional hints
about its focus. After selecting a referenced paper as a candidate for
inclusion, we studied it in detail and evaluated, whether it actually
contributes to broadening the notion of NFRs in DevOps. Specifically, we
examined which properties (cf. Table \ref{tab:data_extraction}, P5-P7) we can extract from a
study and thus how it could contribute to answering the research
questions more thoroughly. We only included papers of which we could
extract at least two properties. Contrarily, we excluded all papers from
being further regarded in the snowballing process if they were either
too generic regarding their NFRs scope, did not provide new insights
into this topic, or because they solely elaborated on basic common
knowledge. Also, we recorded which papers we have already examined and
the rationale for either inclusion or exclusion from snowballing.

Starting with the set of papers retrieved from the structured search we
conducted the backward snowballing process 4 times, resulting in the
following statistics:
\begin{itemize}
    \item \textbf{Start set:} Out of the references of 115 papers originating from the
    structured literature search, we selected 38 candidates of which 9
    papers were included, i.e., efficiency = 9/38 = 23.7\%.
    \item \textbf{Iteration 1:} 26 candidates originating from snowballing from start
    set and 7 papers were included, i.e., efficiency = 7/26 = 22.2\%.
    \item \textbf{Iteration 2:} 35 candidates originating from snowballing from
    iteration 1 and 9 papers were included, i.e., efficiency = 9/35 =
    25.7\%.
    \item \textbf{Iteration 3:} 12 candidates originating from snowballing from
    iteration 2 and 2 papers were included, i.e., efficiency = 2/12 =
    16.7\%.
    \item \textbf{Iteration 4:} 15 candidates were examined from iteration 3, but no
    papers were further included, i.e., efficiency equals to 0\%.
\end{itemize}

We stopped the snowballing procedure after 4 iterations as we did not
find any further papers being relevant for inclusion in the final set.
The overall efficiency of our snowballing procedure, that is calculated
on all candidates, equals to (9+7+9+2+0)/(38+26+35+12+15) = 27/126 =
21.4\%. During the snowballing process we also noticed that each
candidate paper must be analyzed thoroughly, as an analysis spanning
just the title and abstract might often be misleading. Also, we observed
that the additional snowballing procedure in many cases led to relevant
papers of other domains being included, e.g., from the business, product
and strategic management domain, which we would not have without this
step. It shall be noted that actually those papers helped us gaining a
better understanding about the scope and notion, particularly, of
non-engineering-focused NFRs in the software lifecycle.

\subsection{Data extraction}
\label{sec:data_extraction}

In the next step, we extracted the predefined properties shown in Table
\ref{tab:data_extraction} from each selected primary study. This task was done by each
researcher separately using a tool\footnote{MAXQDA Plus, https://www.maxqda.com/} for qualitative text analysis.
Overall, we extracted 2 categories of data that also map with the
research questions: properties P1-P5 focus on the analysis of the
publication population, properties P6-P8 capture the overall focus,
themes, and objectives of the NFRs described in a primary study.

\begin{table}[]
\centering
\begin{tabular}{P{30pt}P{230pt}C{100pt}}
\toprule
\textbf{ID} & \textbf{Property} & \textbf{Research question} \\ \hline
P1 & Research facet & RQ1.1 \\ \hline
P2 & Contribution facet & RQ1.2 \\ \hline
P3 & Publication channel & RQ1.3 \\ \hline
P4 & Publication venue & RQ1.4 \\ \hline
P5 & Publication year & RQ1.4 \\ \hline
P6 & Focus area of NFRs in DevOps & RQ2 \\ \hline
P7 & Themes associated with NFR focus area in DevOps & RQ3 \\ \hline
P8 & Objectives associated with theme of NFR focus area & RQ4 \\
\bottomrule
\end{tabular}
\caption{Data extraction form.}
\label{tab:data_extraction}
\end{table} 

In the following we describe the data extraction for each property from
the primary studies.

\subsubsection{Primary study properties (P1--P5):}

We extracted 4 properties for answering RQ1.1-RQ1.4, which are briefly
described below. The detailed description of the classification schemes
and these properties is given in \ref{appendix:appendix_a}, Table \ref{tab:classification_scheme_facets}.
\begin{enumerate}
    \item \textit{Research facet (P1):} Categorizes the studies by applied research
    design, i.e., how the statements presented in the studies have been
    evaluated. We used existing classification schemes \cite{wieringa_requirements_2005,petersen_systematic_2008} 
    and distinguished between experience reports, evaluation/validation research, philosophical/opinion papers, and solution proposals.
    \item \textit{Contribution facet (P2):} Maps the different outcomes of the
    studies. Again, we followed existing classification schemes \cite{shaw_writing_2003,paternoster_software_2014} and defined 6 categories: model, method, theory, framework, guideline, and lessons learned.
    \item \textit{Publication channel (P3), venue (P4), and year (P5):} Captures the
    publication channels, venues, and years of the studies.
\end{enumerate}

\subsubsection{Focus areas, themes, and objectives of NFRs in DevOps (P6--P8):}

To examine the focus areas and objectives associated with NFRs in
DevOps, we extracted 3 properties from the studies which help answering
RQ2 and RQ3. 
\begin{enumerate}
    \item \textit{General focus area of NFRs in DevOps (P6):} Summarizes the general
    focus of NFRs in DevOps described in the studies. To facilitate the
    inductive coding necessary to capture the qualitative nature of this
    property, we setup 7 different focus areas of NFRs ab initio:
    customer, development, operation, governance, quality, organization,
    and business. An elaborate definition of these focus areas is given
    in \ref{appendix:appendix_a}, Table \ref{tab:classification_scheme_focus_areas}. We derived these focus areas from our pilot
    study when developing the search strings (cf. Section \ref{sec:research_process}) and
    encountered them recurrently among these papers.
    \item \textit{Themes (P7) and objectives (P8) associated with focus areas of NFRs
    in DevOps:} We introduced the concept \textit{"objectives"} to capture the
    underlying goals and motivations tacitly associated with NFRs in
    DevOps and grouped them by spanning theme. In most studies, these
    objectives are either expressed vaguely, outside the research scope
    of the study, or not clearly distinguished from functional
    requirements \cite{broy_rethinking_2015}. If the objective was clearly stated in a
    study, we codified the respective statements. Otherwise we codified
    the origin and context that an NFR was mentioned in or the
    stakeholder(s) who explicated an NFR.
\end{enumerate}

\subsection{Data analysis and interpretation}
\label{sec:data_analysis}
We used descriptive statistics to analyze the data relating to
RQ1.1-RQ1.4. Particularly, we analyzed these data regarding the
frequencies of research and contribution facets, publication channels,
and years. In addition, we used logarithmic regression for trend
analysis of the publication years. The analysis of the respective data
was done by two researchers separately.

For analyzing the qualitative data relating to RQ2 we used thematic
synthesis \cite{cruzes_recommended_2011}. The classification schema for
analyzing the extracted properties was derived from the pilot study
during developing the search strings (cf. Section \ref{sec:research_process}). Thus, this
classification scheme was available right at the beginning of the data
analysis and not developed iteratively. We solely used inductive coding
for analyzing the data of this research question. The main objective of
this analysis was to match the codings, which marked the relevant
passages in the text of the studies, to the given NFR focus areas. We
however revised this classification schema regularly to assure that the
labeling of the NFR focus areas suitably describe the related studies.
The analysis of the studies according to the given classification schema
was done by two researchers separately, followed by a final review of
this classification by all researchers conjointly.

In contrast, for analyzing the data related to RQ3 to identify the
themes and objectives associated with NFRs in DevOps, we used inductive
and deductive coding. First, we labelled the relevant text fragments in
the studies using an inductive approach and derived descriptive codes on
the level of individual studies. Then, after labelling all studies this
manner, we iteratively derived more generic codes that span and suitably
describe the more specific codes. For example, we encountered text
fragments in the studies that reflect on the use of certain versions of
libraries, compare advantages of certain frameworks or elaborate on
specific programming patterns to assure compatibility with a framework.
In this case we synthesized these text fragments through the code
\textit{framework compliance}. After we had developed an initial list of codes,
we revised each code to check whether it suitably matches the content of
the associated study and also that it doesn't leave out important
semantic subtleties of the encoded text fragments. When reviewing this
initial list of codes, we iteratively refined codes and reviewed if the
new codes better fit to the referred text fragments. We repeated this
step multiple times until we finally agreed upon a consolidated list of
codes.

Specifically, when extracting the themes and objectives described in the
studies, we noticed that many studies cannot be attributed to a single
focus area. Instead, they describe multiple focus areas, themes and thus
also different objectives underlying NFRs in DevOps. In these cases, if
relevant for answering the research questions, we separately encoded all
described focus areas, themes, and objectives in a single study.

\section{Overview of the state-state-of-the-art: NFRs in DevOps (RQ1)}
\label{sec:overview_rq1}
After reviewing 229 primary studies retrieved from the academic
databases (cf. Figure \ref{fig:research_process}) and an additional snowballing step, we selected
142 primary studies for this mapping study. In this section we present
our findings from examining the bibliometrics of the study population,
such as the research and contribution facets, used publication channels
and publication frequency using descriptive statistics. The primary
studies were classified according to the classification schemes, as
defined in \ref{appendix:appendix_a}, Tables \ref{tab:classification_scheme_facets} and \ref{tab:classification_scheme_focus_areas}. While in this section we present our
summarized findings, the details how we categorized each primary study
according to these facets are elaborated in \ref{appendix:appendix_b}, Table \ref{tab:systematic map}.

\subsection{Research facets (RQ1.1)}
\begin{figure*}[ht]
\centering
\includegraphics[scale=0.7]{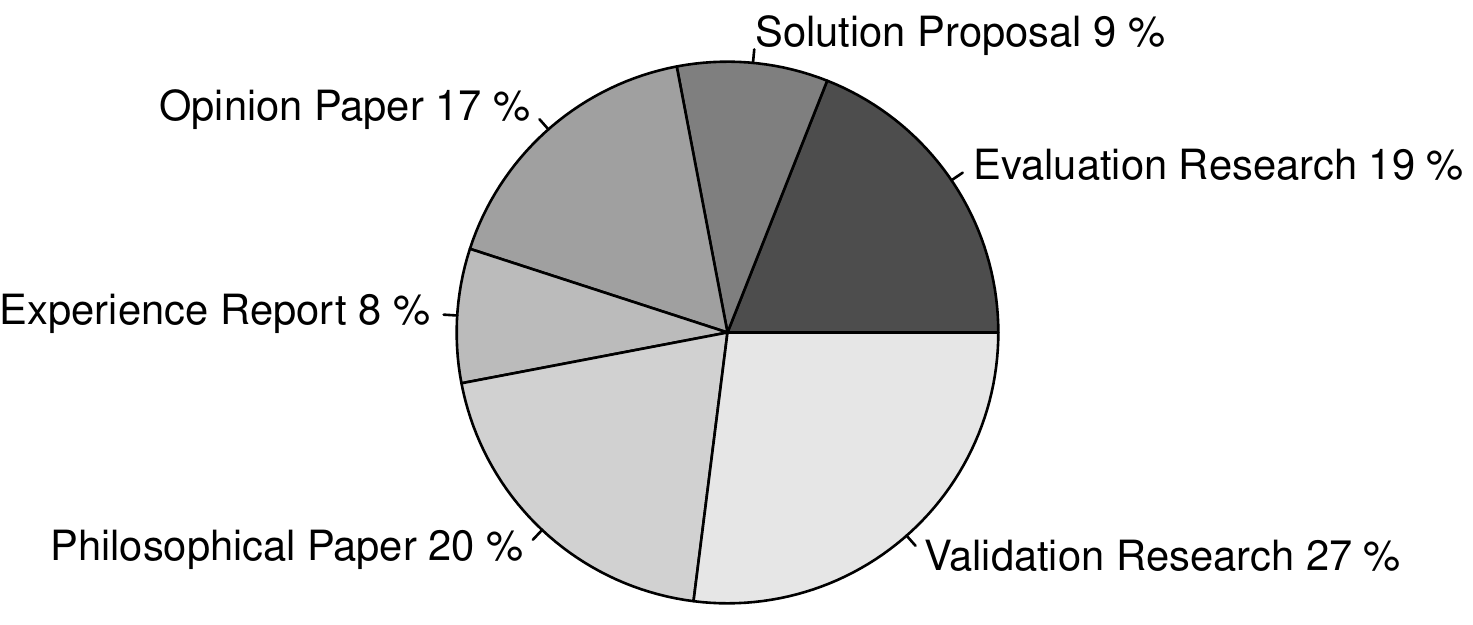}
\caption{Distribution of research facets. \label{fig:research_facets}}
\end{figure*}

Most primary studies on the state-of-the-art handling of NFRs in DevOps
present practical solutions or experiences (63\%, comprising
evaluation/validation research, experience reports, and solution
proposals), contrasted by 37\% of studies being solely theoretical
(represented through philosophical/opinion papers and solution
proposals) expressing the personal opinion of the authors and without
claiming practical applicability. Figure \ref{fig:research_facets} shows the distribution of the
6 research facets. Interestingly, only 8\% of primary studies share
practical experiences regarding handling NFRs in DevOps. An innovative
contribution to this research area can primarily be found by 29\% of the
studies, contributed through philosophical papers and solution
proposals. Contrarily to these both types of works, opinion papers
having a share of 17\% of the studies do not rely on related work or any
sound research methodology and thus also do not claim for applicability,
feasibility or suitability of the presented ideas.

To summarize, practically applicable and empirically validated
approaches for handling NFRs in DevOps only constituted 19\% of the
primary studies, represented by the studies based on evaluation
research. The remaining studies either are not empirically validated, solely
theoretical, or primarily innovate without any form of structured
validation.

\subsection{Contribution facets (RQ1.2)}
\begin{figure*}[ht]
\centering
\includegraphics[scale=0.7]{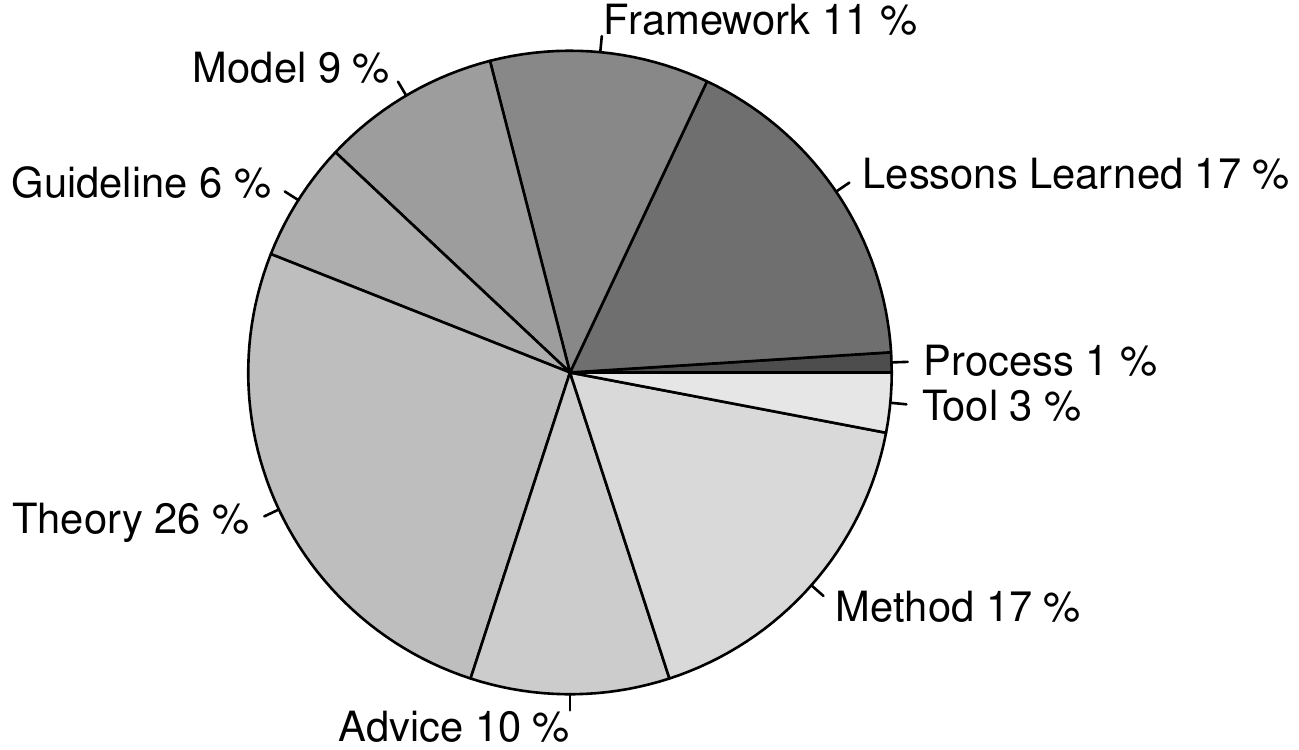}
\caption{Distribution of contribution facets.\label{fig:contribution_facets}}
\end{figure*}

In Figure \ref{fig:contribution_facets} we show the distribution of the contribution facets of the
studies. Many studies contributed to the theory of NFRs in DevOps (26\%,
e.g., requirements prioritization criteria \cite{riegel_systematic_2015},
share experiences and lessons learned in practice (17\%) such as the
capabilities and challenges of DevOps faced in industrial settings
\cite{senapathi_devops_2018}, or describe general advices for handling NFRs
in DevOps (10\%), such as the organizational change associated with
DevOps adoption \cite{dornenburg_path_2018}. On the other side, 40\% of the studies
contribute concrete approaches, comprised of methods (17\%) (e.g., for
evaluating the fulfillment of NFRs towards value streams \cite{saadatmand_toward_2012}, frameworks (11\%), (e.g., for guiding the evaluation of NFRs
specially in DevOps \cite{di_nitto_software_2016}), models (9\%) e.g., for
feature ideation from user interactions \cite{aspara_creating_2013}, and
tools (3\%) for instance to support customer value estimation during
feature ideation \cite{almeida_assessing_2015}.

Processes for handling NFRs in DevOps are contributed by only 1\% of the
studies. It can be summarized that the emphasis of the studies in this
field is on theoretical and conceptual contributions (60\%, comprised of
17\% reports about lessons learned, 6\% guidelines, 26\% theoretical
analyses, 10\% advices and 1\% presenting processes).

\subsection{Publication channels and venues (RQ1.3)}
We also analyzed which channels are used for publishing the studies and
illustrate this distribution in Figure 4. The smallest share of studies
dealing with the handling of NFRs in DevOps is published in books (as
chapters), comprising only 11\% of all studies. The remaining 89\% of
studies are either published on scientific conferences (44\%) or journals
(45\%), which typically follow a stricter review process than
conferences. While the dominance of journal publications is not a valid
measure of the quality of the studies, it might be an indicator of high
research efforts spent in this field.
\begin{figure*}[ht]
\centering
\includegraphics[scale=0.8]{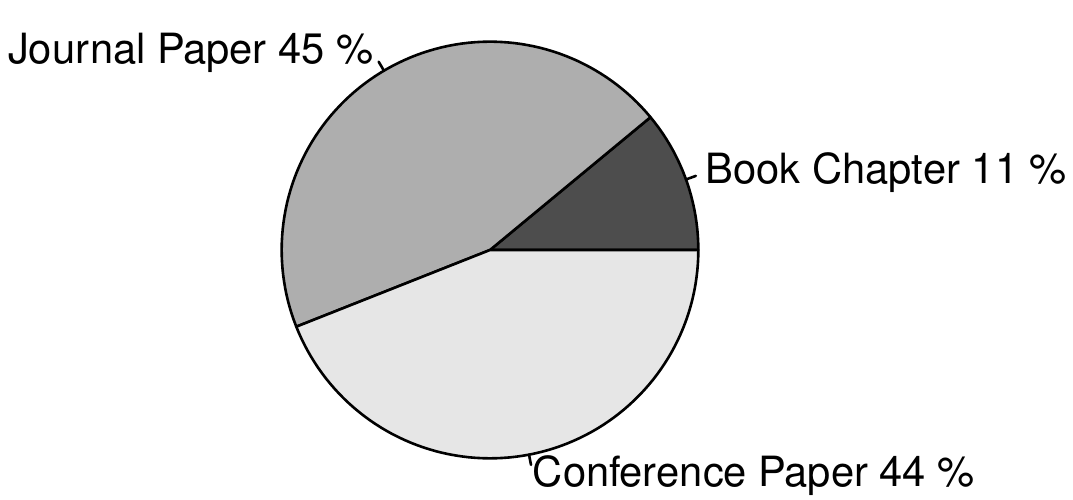}
\caption{Distribution of publication channels.\label{fig:publication_channels}}
\end{figure*}

In addition, we examined the publication venues of the reviewed conference and journal publications, as these channels build the majority of the studies published in this field. In Table \ref{tab:main_venues_and_channels} we outline the channels and their main publication venues, i.e., where more than 3 studies have been published, and the concrete number of studies published there.
\begin{table}[h!]
\centering
\begin{tabular}{P{95pt}P{290pt}C{35pt}}
\toprule
\textbf{Publication channel} & \textbf{Venue} & \textbf{Studies} \\ \hline
\multirow{10}{*}{Conferences} & International Conference on Software Business (ICSOB)  & 6 \\ \cline{2-3}
& International Working Conference on Requirements Engineering: Foundation for Software Quality (REFSQ) & 5 \\ \cline{2-3}
& International Conference on Product-Focused Software Process Improvement (PROFES) & 5 \\ \cline{2-3}
& International Symposium on Empirical Software Engineering and Measurement (ESEM) & 4 \\ \cline{2-3}
& International Conference on Software Engineering (ICSE) & 4 \\ \cline{2-3}
& International Workshop on Quality-Aware DevOps (QUDOS) & 3 \\ \cline{2-3}
& International Conference on Evaluation and Assessment in Software Engineering (EASE) & 3 \\ \cline{2-3}
& International Workshop on Real-Time Business Intelligence and Analytics (BIRTE) & 3 \\ \cline{2-3}
& Euromicro Conference on Software Engineering and Advanced Applications (SEAA) & 3 \\ \cline{2-3}
& International Conference on Agile Software Development (XP) & 3 \\ \midrule
\multirow{5}{*}{Journals} & IEEE Software & 19 \\ \cline{2-3}
 & Journal of Systems and Software & 7 \\ \cline{2-3}
 & Journal of Software: Evolution and Process & 4 \\ \cline{2-3}
 & Empirical Software Engineering & 3 \\ \cline{2-3}
 & Journal of Software Maintenance and Evolution: Research and Practice & 3 \\ \hline
\end{tabular}
\caption{Main publication venues of the studies.}
\label{tab:main_venues_and_channels}
\end{table}
Studies in this field typically are presented on business-oriented (e.g., ICSOB and PROFES), requirements (e.g., REFSQ), and software engineering conferences (e.g., ESEM, EASE, ICSE, and SEAA) or published predominantly in software-focused journals such as "IEEE Software" or "Journal of Systems and Software".

\subsection{Publication frequency over years (RQ1.4)}
\begin{figure*}[ht]
\centering
\includegraphics[scale=0.65]{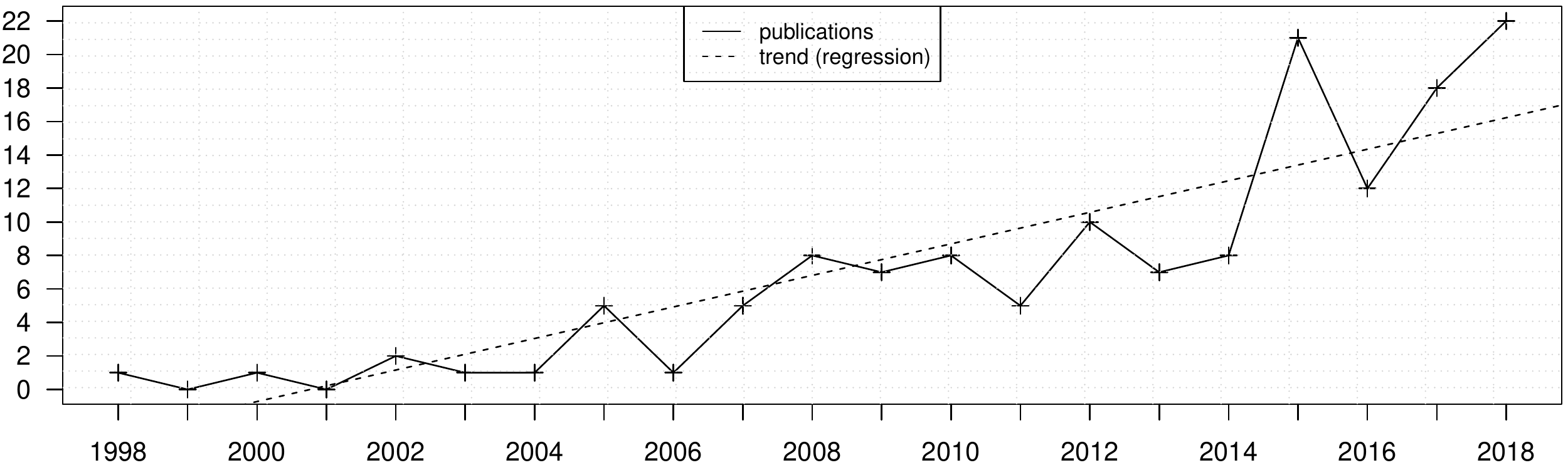}
\caption{Publication distribution by year.\label{fig:year_trend}}
\end{figure*}

As the final step of the quantitative description of the study
population, we also analyzed the number of publications per year. As
depicted in Figure \ref{fig:year_trend}, the reviewed studies were published between 1998
and 2018. The figure also shows the steady incline in publication over
the years, illustrated through the dotted regression line.

Although the first 12 studies (9\%) were published between 1998 and 2006,
rises in the publication frequency can be seen in the years 2007 and
again in 2015. A noticeable and steady publication frequency can be
observed since 2015 where the publication frequency almost doubled
compared to former years. This also indicates that the research field
examining the handling of NFRs in DevOps is still in an early stage,
compared with other fields in software engineering but gaining more
interest in recent years. The continuous rise between 2007 and 2015 can
most probably be attributed to the rising academic interest of NFRs in
software engineering and specially to influencing works in this domain
\cite{cleland-huang_goal-centric_2005,glinz_non-functional_2007,chung_non-functional_2009}. The
second rise in publication frequency since 2015 may be explained with
the advent of the DevOps culture that triggered respective academic
research in this area \cite{smeds_devops:_2015,ebert_devops_2016,jabbari_what_2016}.
\section{Focus areas of NFRs in DevOps (RQ2)}
\label{sec:focus_areas_rq2}
The 142 primary studies cover a multitude of different focus areas – software quality, operational aspects, the structuring of source code to ensure exchangeability of developers, or the mapping of software artefacts to value streams. We categorized the described NFRs in DevOps by 7 focus areas, according to the classification scheme elaborated in \ref{appendix:appendix_a}, Table \ref{tab:classification_scheme_focus_areas}. This classification scheme was derived from the overall focus areas we repeatedly encountered when reviewing the primary studies during piloting the search string (cf. Section \ref{sec:searchstrategy}). In this section we illustrate the dominant focus areas of NFRs in DevOps and the typical research and contribution facets of each focus area. Complementary, we describe the themes and objectives of each focus area along with examples in Section \ref{sec:themes_rq3}.

\subsection{Customer focus area}

\begin{figure*}[ht]
\hspace{-5pt}
\centering
\includegraphics[scale=0.62]{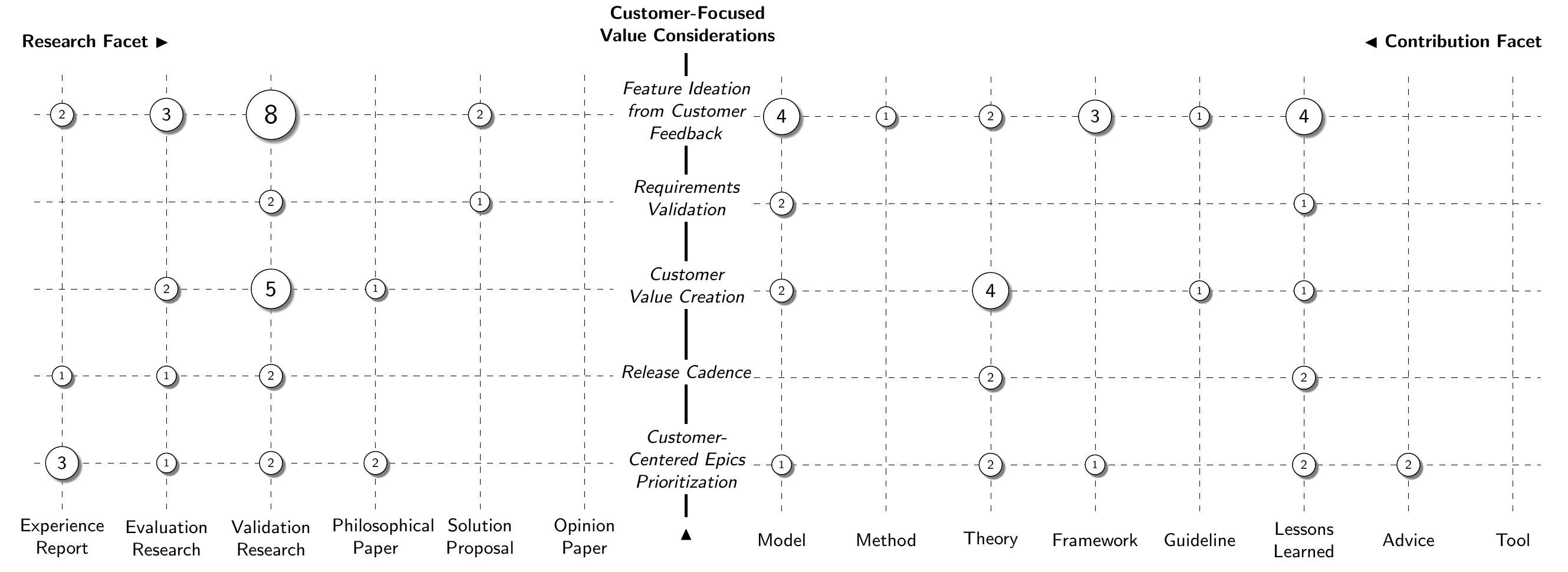}
\caption{Facets and themes of studies examining customer-focused NFRs.\label{fig:customer_focus_area}}
\end{figure*}

We selected 38 studies (27\%) of the study population to be relevant for this focus area, as they elaborate on methods for increasing customer experience, sketching novel features, and monitoring customer-centered improvements throughout the DevOps cycle. In Figure \ref{fig:customer_focus_area} we show the distribution of the research and contribution facets of the studies related to this focus area, along with the 5 main themes recurrently described in the studies.

The majority of studies in this area follows validation research principles, i.e., irrespective their novelty these approaches have not been empirically validated. Only 7 studies in this area are empirically validated (evaluation research facet). However, 6 studies elaborate on concrete practical experiences in the context of customer-focused NFRs in DevOps. These are also supported by 3 theoretical proposals sketching new ideas, mainly for the prioritization and alignment of agile requirements to increase customer experience.
Considering the contributions of the studies, most studies contribute theories and concepts or report on lessons learned either from a software/requirements engineering or product management perspective. Models and frameworks for customer-centricity, supported through NFRs in DevOps, are contributed by 13 of the studies.

\subsection{Development focus area}
\begin{figure*}[ht]
\hspace{-5pt}
\centering
\includegraphics[scale=0.62]{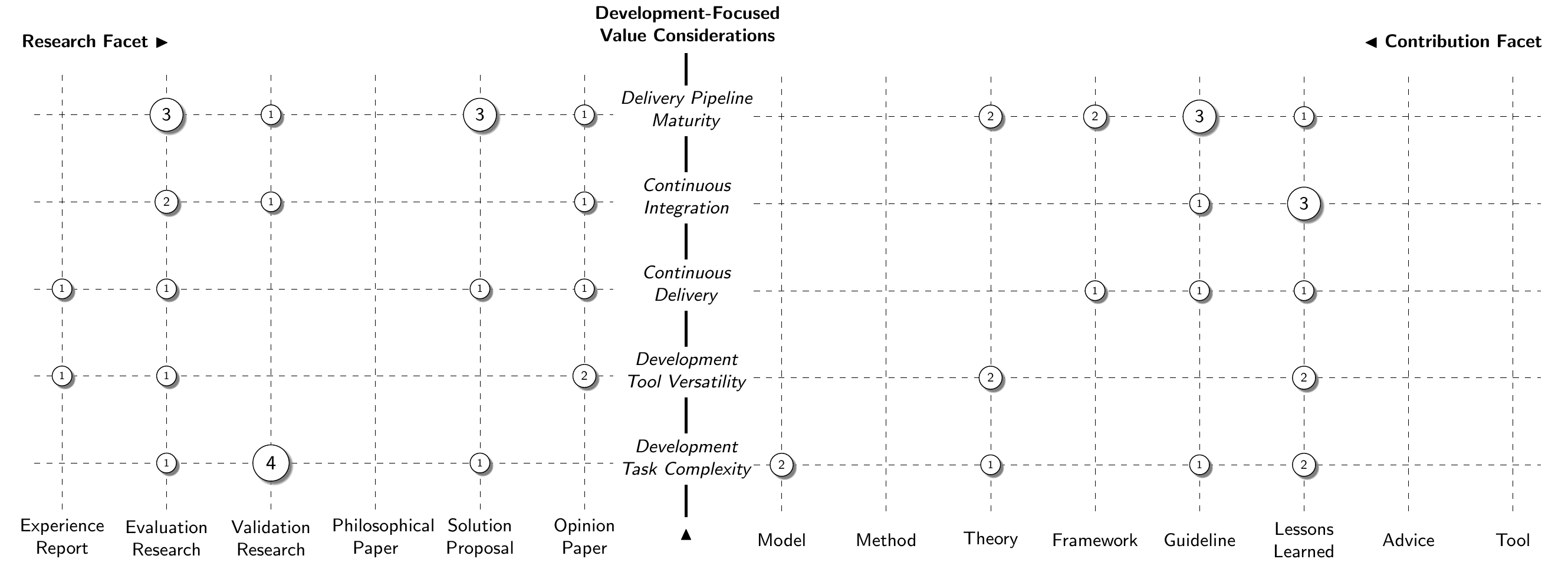}
\caption{Facets and themes of studies examining development-focused NFRs.\label{fig:development_focus_area}}
\end{figure*}

In total, 26 of the 142 studies (18\%) elaborate on NFRs in DevOps that focus on the efficiency of the software development process, e.g., certain restrictions that must be obeyed at an early development or requirements specification stage or particular considerations for continuous integration and delivery. We again picture the distribution of the research and contribution facets as well as the 5 main themes of the respective studies in Figure \ref{fig:development_focus_area}.

Considering the research facets, in this group most studies are empirically (8 studies, evaluation research facet) or theoretically (6 studies, validation research facet) validated. This most likely can be explained through the high availability of data accruing in the development process as well as the availability of experts that can be consulted for empirical studies in this context \cite{martinez-fernandez_towards_2018,shahin_beyond_2017}. Contrarily, 12 studies (46\%, comprised of 2 experience reports, 5 solution proposals, and 5 opinion papers) of the 26 studies focusing on development-centered NFRs have not used any form of validation. 

The majority of 15 studies (58\%) in this focus area contribute practical experiences, comprised of guidelines (6 studies) and lessons learned (9 studies). Concrete frameworks, methods, and models applicable to development-focused NFRs are only contributed by 5 studies (19\%).

\subsection{Operation focus area}
This focus area comprises studies addressing special characteristics and requirements towards the software that must be regarded primarily during its operation. The selected studies of this focus area amount to 20 studies (14\%) of the overall study population and can be split up into 3 prime themes.
\begin{figure*}[ht]
\hspace{-5pt}
\centering
\includegraphics[scale=0.62]{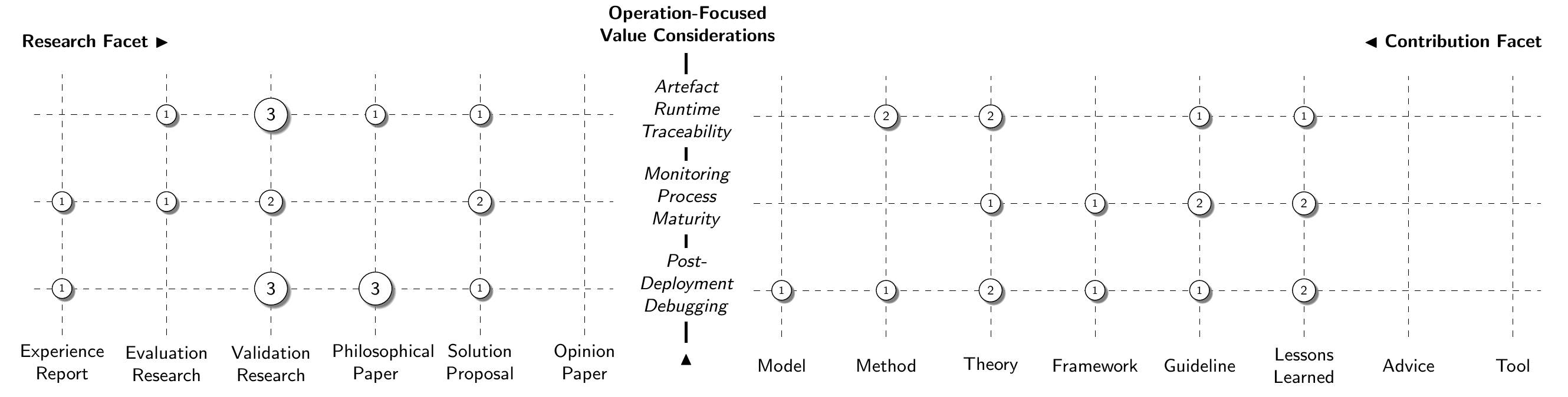}
\caption{Facets and themes of studies examining operation-focused NFRs.\label{fig:operation_focus_area}}
\end{figure*}

Again considering the research facets, only 2 of the 20 studies (10\%) are empirically validated and the benefits and drawbacks clearly elaborated (evaluation research facet). Contrarily, 8 of the 20 studies (40\%) use literature reviews or are based on quantitative case studies (validation research). Compared to these numbers, only 8 of the 20 studies (40\%, comprised of 4 philosophical papers and 4 solution proposals) lack any validation but contribute innovative new ideas to NFRs related to software operation.

A proportionally large share of 9 of 20 studies (45\%) representing this focus area contribute guidelines or report on lessons learned from practical software projects. Concrete approaches for handling software operation-focused NFRs are contributed by 6 studies (30\%, comprised of models, methods, and frameworks). Finally, theoretical inputs to this focus area come from 5 studies (25\%).

\subsection{Governance focus area}
\begin{figure*}[ht]
\hspace{-5pt}
\centering
\includegraphics[scale=0.62]{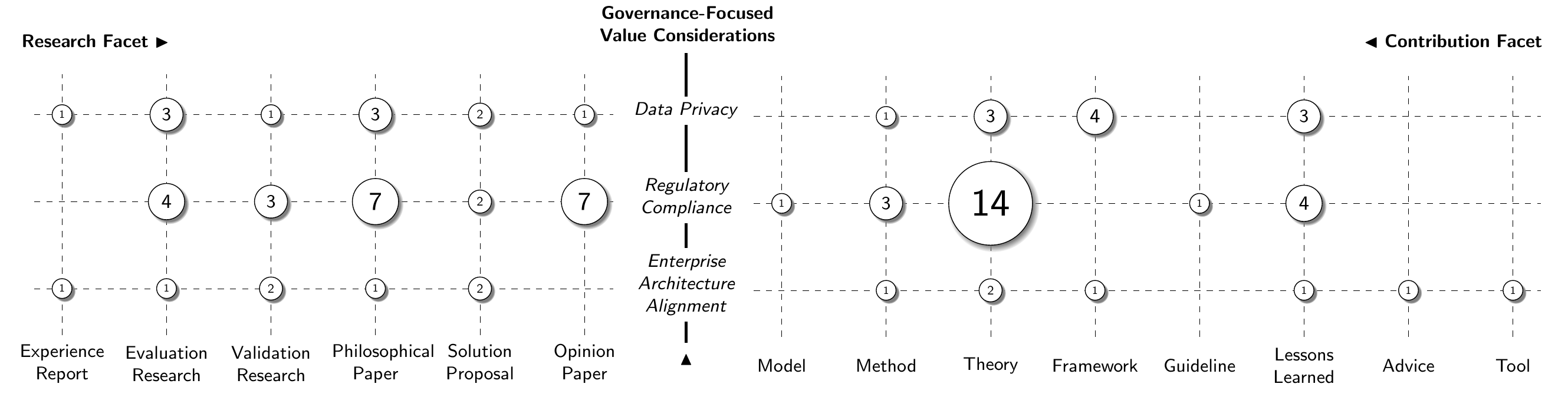}
\caption{Facets and themes of studies examining governance-focused NFRs.\label{fig:governance_focus_area}}
\end{figure*}

Studies elaborating on NFRs that focus on (legal) liabilities, regulatory or technological compliances that must be obeyed and monitored during DevOps are captured by the governance focus area. We selected 41 studies (29\%) being representative for this focus area and extracted 3 main themes linked with the NFRs described in the studies, as illustrated in Figure \ref{fig:governance_focus_area}.
In this focus area, 19 of 36 studies (53\%, comprising 11 philosophical and 8 opinion papers) can be categorized as being solely theoretical and thus dominate among the research facets. Apparent is the small number of experience reports in this focus area, only being attributed to 2 studies (6\%). This is interesting insofar, as the comparatively high number of theoretical papers indicates a certain relevance of the related topics, at least from an academic perspective. Contrarily, practical approaches for handling governance-focused NFRs in DevOps are presented by 14 studies (39\%), comprised of 8 studies being empirically validated (evaluation research facet) and 6 studies presenting literature reviews or quantitative, labor-driven experiments.

The dominance of theoretical works can also be observed among the contribution facets of the works, totaling to 19 studies (53\%). Operational approaches for handling governance-focused NFRs in the DevOps cycle are contributed by 12 studies (33\%, comprised of 1 model, 5 methods, 5 frameworks, and 1 tool described in the studies). Practical guidance for implementing governance-centered NFRs in DevOps is contributed by 10 of the 36 studies (28\%, comprised of 1 guideline, 8 lessons learned, and 1 advice).

\subsection{Technical quality focus area}
\begin{figure*}[ht]
\hspace{-5pt}
\centering
\includegraphics[scale=0.62]{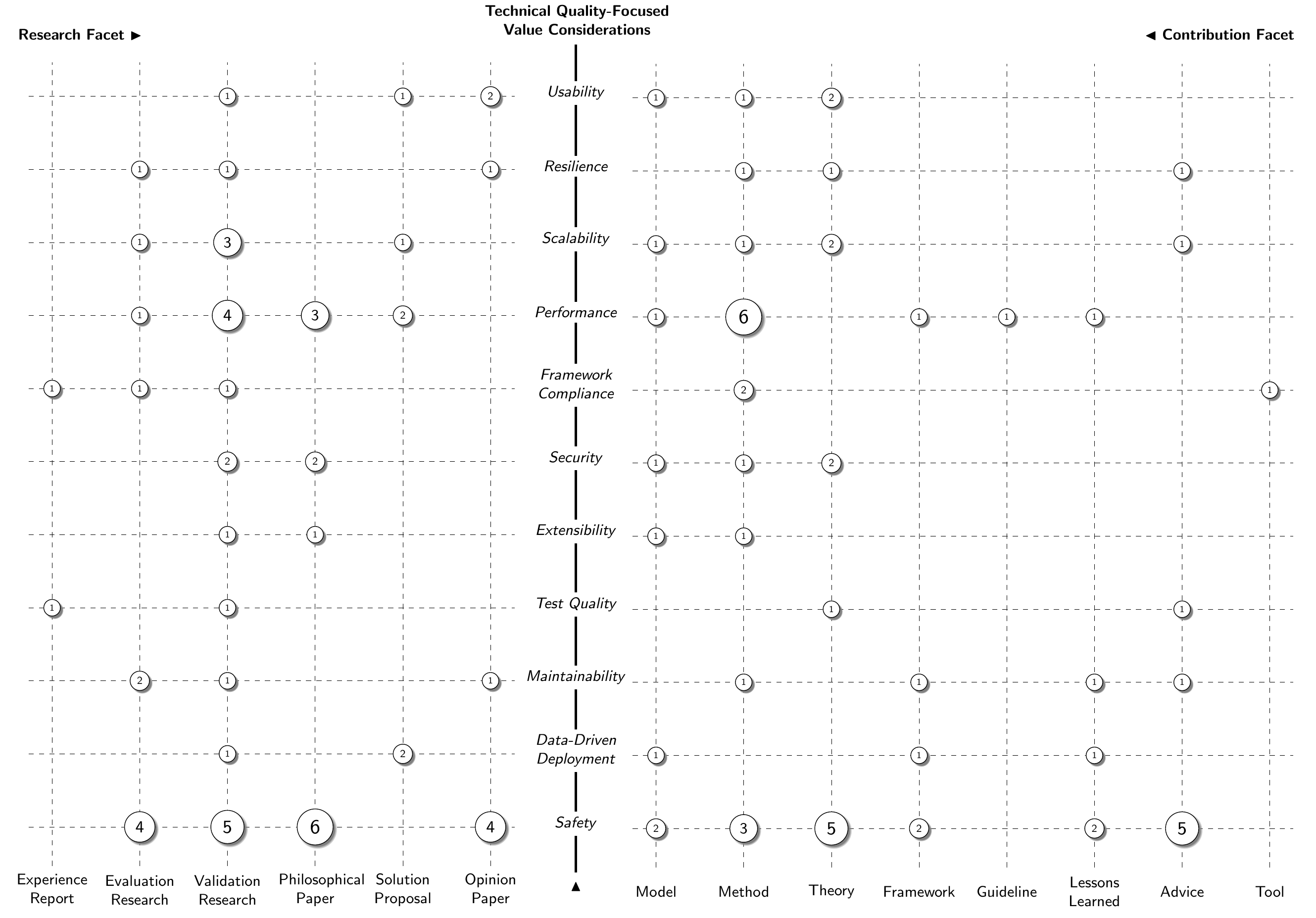}
\caption{Facets and themes of studies examining technical quality-focused NFRs.\label{fig:quality_focus_area}}
\end{figure*}

This focus area addresses the common engineering-focused NFRs of software, ranging from static qualities such as the maintainability of the source code to dynamic qualities such as the response time or memory consumption. Due to the existing classification schemes, such as ISO/IEC 25010 \cite{international_organization_for_standardization_iso/iec_2018}, the themes described in the studies were categorized on a much finer level than of the other focus areas. Figure \ref{fig:quality_focus_area} illustrates the primarily engineering-oriented themes of this focus area.
Also, due to the broad coverage of this type of NFRs in academic literature, the majority of studies retrieved from the databases are applicable to this focus area. Overall, we selected 59 studies (42\%) for this focus area and extracted 11 representative themes. With 21 respective studies (36\%), validation research takes the greatest share among the research facets, followed by 10 studies (17\%) applying evaluation research methodologies. Compared with the 47\% of studies attributed to other research facets (comprised of 12 philosophical, 6 solution proposals, 8 opinion papers, and 2 experience reports) the emphasis in this focus area apparently is on validated research. The majority of research in this focus area is conducted theoretically through quantitative testing methods or conducting experiments with prototypes. These studies nonetheless lack a clear elaboration on the specific benefits and limitations of their approaches in practical contexts. However, a very high number of studies applied empirical methods additionally. Specially, these studies examined the use of certain measurement or evaluation techniques of NFRs in practice or tested novel approaches in practical settings and industrial contexts followed by e.g., expert interviews from different domains. Also interesting is the comparatively high number of philosophical papers, which e.g., sketch possible ways of intertwining functional and non-functional requirements \cite{yang_linking_2012} or challenge the scope and notion of NFRs and their demarcation from functional requirements \cite{broy_rethinking_2015,eckhardt_are_2016}. 

We found 17 studies (29\%) contributing concrete methods for handling quality-related NFRs in DevOps, complemented by 8 studies (14\%) describing suitable models and 5 studies (8\%) presenting frameworks for that purpose. Practical experiences for tackling software quality particularly in DevOps are contributed by 9 studies (15\%) in the form of concrete advices and by 4 studies (7\%) elaborating lessons learned in handling quality-related NFRs in DevOps. Theoretical contributions such as e.g., classification schemes come from 13 of the 59 studies (22\%). 

\subsection{Organization focus area}
\begin{figure*}[ht]
\hspace{-5pt}
\centering
\includegraphics[scale=0.62]{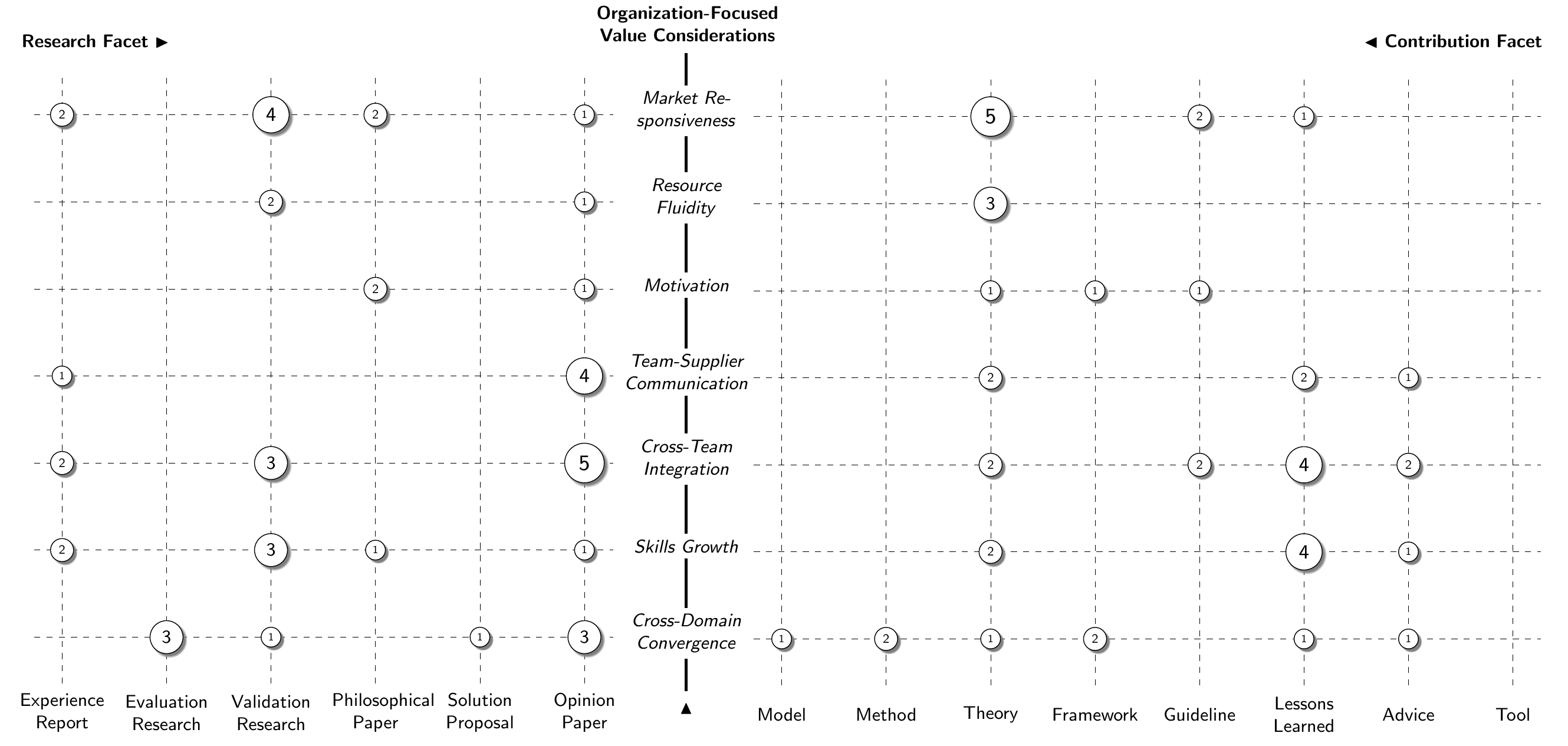}
\caption{Facets and themes of studies examining organization-focused NFRs.\label{fig:organization_focus_area}}
\end{figure*}

Studies examining NFRs in DevOps that target the organization level within companies are summarized in this focus area. We selected 45 of the 142 studies (32\%) to be applicable for this focus area and subdivided their described NFRs into 7 themes, as illustrated in Figure \ref{fig:organization_focus_area}. Implicitly, the NFRs of this focus area address the organizational capabilities and properties that must be in place to effectively implement business-driven themes with software in a company. Thus, they address human factors such as motivation or the required support for skills growth as well as team/supplier communication effectiveness. In contrast to other focus areas, the respective NFRs can mainly be expressed and evaluated on a qualitative basis.
We categorized 21 studies (47\%, comprised of 16 opinion and 5 philosophical papers) as theoretical research works without empirical validation. On the contrary side, 16 studies (36\%, comprised of 13 validation and 3 evaluation research works) presented validated research, mainly through systematic or narrative literature reviews. Due to the applied research methods, these studies do not claim for practical applicability of their findings. A small number of 3 studies applied research methods that also allow to examine specific benefits of their approaches, e.g., through conducting expert interviews or case study research. Lastly, 8 studies (18\%, comprising 1 solution proposal and 7 experience reports) examine the handling of organization-focused NFRs in practice and report about possible ways for improvement and practical challenges, without sketching a generally applicable and validated methodology.

In total, 16 studies (13\%) in this focus area contribute theoretical concepts such as success factors for agile transformations \cite{dikert_challenges_2016}, integrations of the Scaled Agile framework (SAFe) for software development \cite{laanti_characteristics_2014,putta_benefits_2018} or differences between quality management strategies \cite{tontini_integrating_2007} on organizational level. Concrete guidance for implementing NFRs in DevOps that address organization-wide themes is contributed by 22 studies (49\%, comprised of 5 guidelines, 12 studies elaborating on lessons learned, and 5 giving advices). These contributions primarily originate from multi-case studies focusing on DevOps capabilities and picture the benefits and challenges faced by the companies \cite{eric_lean_2017,senapathi_devops_2018}. Practically applicable artefacts, mainly complemented by empirical validations in the context of case studies, are contributed by 6 studies (13\%, comprising 1 model, 2 methods, and 3 frameworks).

\subsection{Business focus area}
Finally, the business focus area covers all NFRs that address strategic, business, and cost/revenue considerations that manifest in software NFRs and need to be considered throughout the DevOps cycle. As illustrated in Figure 12, we selected 39 studies (27\%) for this focus area and carved out 5 main themes behind the respective NFRs. The typical themes of this focus area are closely linked with elements of software business models \cite{osterwalder_business_2010,wirtz_business_2011} of the company. The fulfillment of these NFRs is mainly evaluated through dynamic measures.
\begin{figure*}[ht]
\hspace{-5pt}
\centering
\includegraphics[scale=0.62]{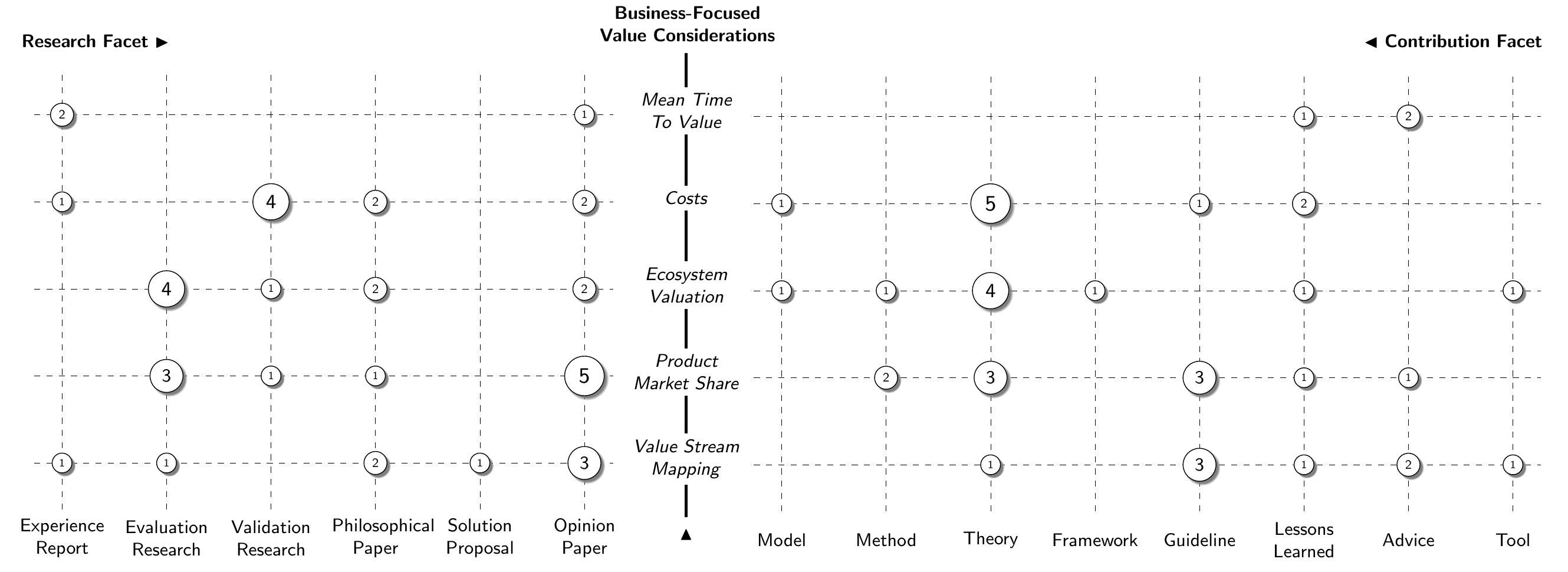}
\caption{Facets and themes of studies examining business-focused NFRs.\label{fig:business_focus_area}}
\end{figure*}
The largest portion of studies in this focus area, totaling 20 studies (51\%, comprising 7 philosophical and 13 opinion papers), deal with business-focused NFRs in DevOps from a solely theoretical perspective, e.g., to outline business models specially for software \cite{fricker_software_2012}, reflect on the suitability of ROI indicators for software \cite{boehm_value-based_2003}, or sketch cost and value estimation techniques \cite{kersten_what_2018}. These theoretical and often biased works express personal opinions or future research directions and are complemented by 14 studies (36\%) following sound validation methods. These studies have either evaluated a concrete methodology in practice (8 studies, evaluation research facet) and elaborate on their benefits and limitations, e.g., to guide value estimation in software projects \cite{almeida_assessing_2015} or conducted literature reviews in any form (6 studies, validation research facet) particularly about business models and value creation from software \cite{aspara_creating_2013,luoma_current_2012}. Lastly, 5 studies (13\%) elaborate on practical experiences (4 studies) and implementation solutions (1 study) for handling business-focused NFRs in DevOps. Particularly these studies focus on the monitoring of the respective NFRs and how to map business-level objectives to software NFRs.

Considering the contribution facets of the studies, there is an emphasis on theoretical works (13 studies, 33\%) examining representative performance indicators that can be applied to software products. Research artefacts that facilitate the practical handling of business-focused NFRs in DevOps, e.g., through methods and tools for deriving quantitative software performance indicators for software products, are contributed by 8 studies (21\%, comprised of 2 models, 3 methods, 1 framework and 2 tools). Best-practices from own or observed experiences of the authors and implementation guidelines are presented by 18 studies (46\%, comprising 7 studies presenting guidelines, 6 studies reporting on lessons learned and 5 studies giving concrete advices).

\section{Themes and objectives of NFRs in DevOps (RQ3)}
\label{sec:themes_rq3}
In this section we describe the recurrent themes of NFRs in DevOps that we identified in the studies and the typical objectives that are linked with them. We group this description by the focus areas presented in the previous section.
\subsection{Themes and objectives behind NFRs of the customer focus area}
\label{sec:customer_themes}
The customer-focused themes of NFRs in DevOps target on optimizing the contributed value to the customer, as illustrated in Table \ref{tab:customer_themes}. An emphasis is on monitoring and analyzing user interaction with the software for data-driven software improvement and shortening of release cycles.

\begin{longtable}{P{190pt}P{62pt}P{185pt}}
\toprule
\textbf{Theme} & \textbf{Frequency} & \textbf{Primary studies} \\ \hline
Feature ideation from customer feedback & \centering 15 & \cite{bjarnason_challenges_2014,bosch_esao:_2014,dzamashvili_impact_2010,erich_qualitative_2017,fabijan_early_2015,fabijan_customer_2015,fagerholm_building_2014,fagerholm_right_2017,forsgren_components_2018,martinez-fernandez_towards_2018,olsson_towards_2015,rodriguez_continuous_2017,sauvola_towards_2015,vanhala_role_2014,yaman_customer_2016} \\ \hline
Requirements validation & \centering 3 & \cite{cao_agile_2008,dzamashvili_impact_2010,olsson_requirements_2016} \\ \hline
Customer value creation & \centering 8 & \cite{carrillo_metrics_2012,drutsa_practical_2015,huang_technology_2013,kevic_characterizing_2017,mehmood_evaluating_2009,parker_platform_2017,riegel_systematic_2015,rodriguez_evaluation_2016} \\ \hline
Release cadence & \centering 4 & \cite{putta_benefits_2018,riungu-kalliosaari_devops_2016,rodriguez_continuous_2017,senapathi_devops_2018} \\ \hline
Customer-centered epics prioritization & \centering 8 & \cite{dingsoyr_exploring_2018,dornenburg_path_2018,erich_qualitative_2017,favaro_managing_2002,martinez-fernandez_quality_2018,riegel_systematic_2015,senapathi_devops_2018,vahaniitty_towards_2008} \\ \bottomrule
\caption{Themes and objectives of customer-focused NFRs.}
\label{tab:customer_themes}
\end{longtable}

\begin{itemize}
	\item \textbf{Feature ideation from customer feedback:} Gathering continuous feedback from the customer for software improvement as well as the ideation of new features. The focus of the improvement mainly is on the functional suitability of the software. For feature ideation, quantitative user interaction data (from user monitoring) is complemented with qualitative data from user surveys.
	\item \textbf{Requirements validation:} Validating the relevance of user requirements throughout the development cycle, e.g., through frequent validation cycles with customers.
	\item \textbf{Customer value creation:} Estimating the contributed value of software throughout the development cycle, e.g., through a \textit{"customer needs scorecard"} \cite{huang_technology_2013} or the analysis of user queries \cite{parker_platform_2017}.
	\item \textbf{Release cadence:} Shortening the time between planning and provisioning of software changes.
	\item \textbf{Customer-centered epics prioritization:} Prioritizing epics and user stories by contributed value of the software change for the user. Contributed user value is estimated e.g., through goal-pursuit time from monitored user interaction \cite{vahaniitty_towards_2008}.
\end{itemize}

\subsection{Themes and objectives behind NFRs of the development focus area}
\label{sec:development_themes}
The themes associated with development-focused NFRs in DevOps target on reducing failure rates in the software development lifecycle, mostly through frequent feedback loops and efficient tool support. We illustrate these themes in Table \ref{tab:development_themes} and the objectives sequentially.

\begin{longtable}{P{190pt}P{62pt}P{185pt}}
\toprule
\textbf{Theme} & \textbf{Frequency} & \textbf{Primary studies} \\ \hline
Delivery pipeline maturity & \centering 8 & \cite{di_nitto_software_2016,forsgren_devops_2018,jabbari_what_2016,kersten_mining_2018,riungu-kalliosaari_devops_2016,rodriguez_continuous_2017,shahin_beyond_2017,vargas_enabling_2018} \\ \hline
Continuous integration &  \centering 4	& \cite{cao_agile_2008,ebert_scaling_2017,kevic_characterizing_2017,shahin_beyond_2017} \\ \hline
Continuous delivery	& \centering 4 & \cite{di_nitto_software_2016,ebert_scaling_2017,forsgren_software_2018,shahin_beyond_2017} \\ \hline
Development tool versatility &	\centering 4 & \cite{ebert_scaling_2017,gall_software_2014,jabbari_what_2016,laanti_characteristics_2014} \\ \hline
Development task complexity	& \centering 6 & \cite{cao_agile_2008,dzamashvili_impact_2010,karlsson_requirements_2007,kersten_what_2018,kersten_mining_2018,kopczynska_empirical_2018,martinez-fernandez_quality_2018} \\ \bottomrule
\caption{Themes and objectives of development-focused NFRs.}
\label{tab:development_themes}
\end{longtable}

\begin{itemize}
	\item \textbf{Delivery pipeline maturity:} Closing feedback loops to facilitate actionable analytics of the data accruing on the different systems in DevOps. The degree of actionability is assessed through concrete metrics and reflects the maturity of the delivery pipeline.
	\item \textbf{Continuous integration:} Assuring correct system, software, and service behavior through continuous integration of smaller code changes into the codebase of the whole product.
	\item \textbf{Continuous delivery:} Reducing the failure rate after software changes through deploying the software in short cycles. Also, the aim is on gaining understanding of the impacts of the software change on operational characteristics and improve architectural decisions thereof \cite{di_nitto_software_2016}.
	\item \textbf{Development tool versatility:} Assuring the suitability and flexibility of the development environments (e.g., IDEs, source code repositories, issue trackers) to support engineers in their development activities and the tracking between source code and compiled software on different systems.
	\item \textbf{Development task complexity:} Reducing complexity of development tasks, e.g., through architectural and code blueprints, best practices, and guidelines for development tasks.
\end{itemize}

\subsection{Themes and objectives behind NFRs of the operation focus area}
\label{sec:operation_themes}
The main objective of the themes linked with operation-focused NFRs in DevOps address the monitoring, debugging, and tracing of runtime artefacts, as illustrated in Table \ref{tab:operation_themes}.

\begin{longtable}{P{190pt}P{62pt}P{185pt}}
\toprule
\textbf{Theme} & \textbf{Frequency} & \textbf{Primary studies} \\ \hline
Artefact runtime traceability & \centering 6 & \cite{cao_agile_2008,cito_context-based_2017,kersten_mining_2018,lian_mining_2017,niu_requirements_2018,riegel_systematic_2015} \\ \hline
Monitoring process maturity & \centering 6 & \cite{di_nitto_software_2016,dikert_challenges_2016,kersten_mining_2018,riungu-kalliosaari_devops_2016,senapathi_devops_2018,shahin_beyond_2017} \\ \hline
Post-deployment debugging &	\centering 8 & \cite{bjarnason_challenges_2014,erich_qualitative_2017,guzman_how_2017,haleem_impact_2015,luoma_current_2012,martinez-fernandez_towards_2018,niu_requirements_2018,vanhala_role_2014} \\ \bottomrule
\caption{Themes and objectives of operation-focused NFRs.}
\label{tab:operation_themes}
\end{longtable}

\begin{itemize}
	\item \textbf{Artefact runtime traceability:} Establishing links between source code artefacts and runtime entities \cite{cito_context-based_2017}. Specially, this needs to regard different versions of the software deployed on different systems and allow to relate monitored operational data back to the originating source code version.
	\item \textbf{Monitoring process maturity:} Effectively and efficiently monitor the software in operation and automatize the integration of the monitoring data for software analytics \cite{riungu-kalliosaari_devops_2016}.
	\item \textbf{Post-deployment debugging:} Facilitating debugging of the operational software, e.g., upon deviations between expected and actual software behavior. Particularly in system-of-system architectures and transactional systems this may lead to specific challenges arising from different network protocols/architectures, and firewalls \cite{erich_qualitative_2017}.
\end{itemize}

\subsection{Themes and objectives behind NFRs of the governance focus area}
\label{sec:governance_themes}
In this focus area, the objectives of the themes behind the NFRs are on assuring data privacy, enterprise architecture, and regulatory compliance. We picture the themes and their frequency in the studies in Table \ref{tab:governance_themes} and the respective objectives afterwards.

\begin{longtable}{P{190pt}P{62pt}P{185pt}}
\toprule
\textbf{Theme} & \textbf{Frequency} & \textbf{Primary studies} \\ \hline
Data privacy & \centering 10 & \cite{bosch-sijtsema_user_2015,di_nitto_software_2016,franch_data-driven_2018,gall_software_2014,ivanov_implementation_2018,kersten_what_2018,robinson_roadmap_2010,rodriguez_continuous_2017,shah_novel_2016,vierhauser_requirements_2016} \\ \hline
Liabilities and regulatory compliance & \centering 19 & \cite{adner_wide_2013,bjarnason_challenges_2014,boudreau_how_2009,dikert_challenges_2016,ebert_scaling_2017,fahssi_enhanced_2015,fricker_software_2012,glinz_non-functional_2007,kersten_what_2018,kevic_characterizing_2017,kittlaus_software_2012,lian_mining_2017,parker_platform_2017,riegel_systematic_2015,rodriguez_continuous_2017,schmeling_composing_2011,timmers_business_1998,zhang_data-driven_2017,zowghi_requirements_2005} \\ \hline
Enterprise architecture alignment &	\centering 7 & \cite{ameller_dealing_2010,fleischmann_coherent_2010,hiisila_combining_2015,martinez-fernandez_towards_2018,shah_novel_2016,svee_consumer_2012,urbinati_role_2018} \\ \bottomrule
\caption{Themes and objectives of governance-focused NFRs.}
\label{tab:governance_themes}
\end{longtable}

\begin{itemize}
	\item \textbf{Data privacy:} Assuring privacy of the data accruing on the different systems, comprised of e.g., interaction traces and  processed sensitive data from the user \cite{franch_data-driven_2018}.
	\item \textbf{Liabilities and regulatory compliance:} Addressing how legal obligations and duties can be implemented by the software, such as the processing of licensing data \cite{fricker_software_2012}, the recording of presented information to customers in financial advisory processes \cite{riegel_systematic_2015}, the calculation scheme for financial ratings of customers, or the processing of data for identifying customers in accounting applications \cite{schmeling_composing_2011}.
	\item \textbf{Enterprise architecture alignment:} Aligning and integrating the software into the enterprise application landscape through architectural guidelines and frameworks, e.g., TOGAF \cite{haren_togaf_2011} or ATAM \cite{kazman_atam:_2000}.
\end{itemize}

\subsection{Themes and objectives behind NFRs of the technical quality focus area}
\label{sec:quality_themes}
The themes in the quality focus area mainly reflect common static and dynamic software quality factors, as also defined in e.g., the ISO/IEC 25010 standard \cite{international_organization_for_standardization_iso/iec_2018}. However, as shown in Table \ref{tab:quality_themes}, the frequency of studies examining certain themes varies among the studies which indicates that they are differently important in DevOps. Due to the apparent similarities of the themes to the ISO 25010 standard, we primarily outline important differences or additions of the respective NFRs in DevOps.
\newline \\
\begin{longtable}{P{190pt}P{62pt}P{185pt}}
\toprule
\textbf{Theme} & \textbf{Frequency} & \textbf{Primary studies} \\ \hline
Usability & \centering 4 & \cite{gonzalez_calleros_towards_2014,navarre_icos:_2009,rich_building_2009,rodriguez_evaluation_2016} \\ \hline
Resilience & \centering 3 & \cite{griffith_industry_2017,lamsweerde_handling_2000,martinez-fernandez_quality_2018} \\ \hline
Scalability & \centering 5 & \cite{ameller_dealing_2010,dingsoyr_exploring_2018,fagerholm_building_2014,putta_benefits_2018,saadatmand_toward_2012} \\ \hline
Performance & \centering 10 & \cite{bosch_introducing_2011,chung_non-functional_2009,cleland-huang_goal-centric_2005,forbrig_rapid_2014,glinz_risk-based_2008,klas_cqml_2009,liu_integrating_2010,matsumotoa_method_2017,robinson_requirements_2006,schmeling_composing_2011} \\ \hline
Framework compliance & \centering 3 & \cite{hecht_tracking_2015,kaiya_quality_2011,r._plosch_collecting_2010} \\ \hline
Security & \centering 4 & \cite{carrillo_metrics_2012,eckhardt_how_2015,fabijan_customer_2015,schief_business_2014} \\ \hline
Extensibility & \centering 2 & \cite{mehmood_evaluating_2009,yang_linking_2012} \\ \hline
Test quality & \centering 2 & \cite{begel_analyze_2014,yang_safety_2017} \\ \hline
Maintainability & \centering 4 & \cite{bosch_speed_2016,lochmann_are_2013,olsson_holmstrom_ad_2017,siavvas_qatch_2017} \\ \hline
Data-driven deployment & \centering 3 & \cite{olsson_requirements_2016,olsson_towards_2015,sauvola_towards_2015} \\ \hline
Safety & \centering 19 & \cite{ahmad_modeling_2015,ameller_dealing_2019,berander_requirements_2005,blaine_software_2008,broy_rethinking_2015,ebert_looking_2015,eckhardt_are_2016,fagerholm_right_2017,herrmann_exploring_2007,leber_value_2014,liu_extension_2009,ozkaya_making_2008,robinson_monitoring_2002,rolland_modeling_2005,vickers_satisfying_2007,vierhauser_requirements_2016,vierhauser_developing_2015,wagner_operationalised_2015,zowghi_requirements_2005} \\ 
\bottomrule
\caption{Themes and objectives of technical quality-focused NFRs.}
\label{tab:quality_themes}
\end{longtable}

\begin{itemize}
	\item \textbf{Usability:} Reducing the time and cognitive load to accomplish goals with software. Task-based methods allow concrete quantification and automatized evaluation of usability \cite{rich_building_2009,rodriguez_evaluation_2016}.
 	\item \textbf{Resilience:} Supporting flexible adaptions of software to ensure product robustness and particularly to diminish remediation costs in DevOps \cite{griffith_industry_2017}.
	\item \textbf{Scalability:} Facilitating automatized deployments to dynamically react on usage peaks. In the DevOps context this involves inserting probes into the source code prior to deployment that can trigger automatic deployments thus ensuring elasticity of the system \cite{ameller_dealing_2010}.
	\item \textbf{Performance:} Providing results to the user of the software in a responsive fashion independently on whether the provided software is a graphical user interface (GUI) or a service endpoint (API). In addition, performance requirements shall be explicated and evaluated on the level of individual features, balanced with their strategic and economic value \cite{berander_requirements_2005,bosch_introducing_2011}.
	\item \textbf{Framework compliance:} Reducing complexity, training efforts, and maintenance costs by adhering to predetermined and evaluated frameworks. For this purpose, most companies use library blueprints or ready-to-use build profiles, which can be integrated with little effort in the DevOps toolchain.
	\item \textbf{Security:} Specially with the proliferation of software ecosystems shared between multiple companies, end-to-end encryption of entered user data and cascaded cryptographic blocks of the user journey are gaining more attention. Most of these security requirements need to be fulfilled on feature-level and require operational support for acquiring the measures in a DevOps context.
	\item \textbf{Extensibility:} Facilitating modification of the software, e.g., through separation and decomposition of functional blocks into self-contained units of work. In DevOps, apart from static code metrics, also functional and latent dependencies between software components can be traced. This can for instance be achieved through labeling software components and extracting cause-effect-relationships from operational data.
	\item \textbf{Test quality:} Assessing test processes, methods, sources, and tools used for testing in DevOps, i.e., test strategies, the degree of automation, test coverage, test report dissemination and handling of failed test cases.
	\item \textbf{Maintainability:} Lowering maintenance costs through adhering e.g., to design best practices, coding styles and the early elimination of \textit{code bad smells}. Due to the availability of operational data in DevOps, maintainability requirements shall be balanced with actual feature usage \cite{lochmann_are_2013}.
	\item \textbf{Data-driven deployment:} Evolving development practices to the point where they could continuously deploy and validate individual features, rather than preparing larger product releases. Apparently, this requires continuous collection of post-deployment data for quantitative analysis of this data. These data allow to derive development activities and suitable indicators of deployment efficiency \cite{olsson_towards_2015}.
	\item \textbf{Safety:} Giving guarantees for certain quality attributes under predefined conditions. Specially with the proliferation of software-centric devices in new areas such as autonomous driving or drones, the monitoring of safety quality attributes in a DevOps context gains increasing industry relevance \cite{ebert_looking_2015}.
\end{itemize}

\subsection{Themes and objectives behind NFRs of the organization focus area}
\label{sec:organization_themes}
The organization-focused themes of NFRs in DevOps target on ensuring organizational agility through harmonizing team, supplier, and market requirements. An important and recurrent center point among the themes is the human factor and its influence on software development and operation, as illustrated in Table \ref{tab:organization_themes}.

\begin{longtable}{P{190pt}P{62pt}P{185pt}}
\toprule
\textbf{Theme} & \textbf{Frequency} & \textbf{Primary studies} \\ \hline
Market responsiveness & \centering 8 & \cite{allix_androzoo:_2016,bosch_integration_2010,bosch_introducing_2011,dikert_challenges_2016,forsgren_measuring_2018,laanti_characteristics_2014,putta_benefits_2018,raffo_integrating_2010,yaman_customer_2016} \\ \hline
Resource fluidity & \centering 3 & \cite{foss_fifteen_2017,laanti_characteristics_2014,putta_benefits_2018} \\ \hline
Motivation & \centering 3 & \cite{bosch-sijtsema_user_2015,boudreau_how_2009,west_leveraging_2014} \\ \hline
Team-supplier communication & \centering 5 & \cite{ebert_scaling_2017,fricker_software_2012,kittlaus_software_2012,lindberg_design_2011,senapathi_devops_2018} \\ \hline
Cross-team integration & \centering 10 & \cite{boehm_value-based_2003,bosch_speed_2016,bosch_introducing_2011,dornenburg_path_2018,ebert_scaling_2017,erich_qualitative_2017,forsgren_measuring_2018,forsgren_software_2018,putta_benefits_2018,wolbling_design_2012} \\ \hline
Skills growth & \centering 7 & \cite{berander_requirements_2005,dikert_challenges_2016,ebert_software_2014,erich_qualitative_2017,ozkaya_making_2008,riungu-kalliosaari_devops_2016,senapathi_devops_2018} \\ \hline
Cross-domain convergence & \centering 7 & \cite{ebert_scaling_2017,fabijan_customer_2015,olsson_holmstrom_ad_2017,regnell_supporting_2008,robinson_roadmap_2010,tontini_integrating_2007} \\ 
\bottomrule
\caption{Themes and objectives of technical organization-focused NFRs.}
\label{tab:organization_themes}
\end{longtable}

\begin{itemize}
	\item \textbf{Market responsiveness:} Decreasing the time to respond to customer requests by concretizing them on the level of software functionality. This can be achieved in DevOps e.g., through extracting and mapping operational software characteristics to metrics for Quality Function Deployment (QFD) \cite{raffo_integrating_2010}.
	\item \textbf{Resource fluidity:} Applying software engineering methods, best practices, and standardized tools that allow to shuffle work between the organization when priorities change.
	\item \textbf{Motivation:} Leveraging internal sources for software improvement \linebreak through providing feedback directly and frequently to a certain engineer. In DevOps this can be accomplished for instance through labelling software artefacts and analyzing the commit history automatically.
	\item \textbf{Team-supplier communication:} Empowering teams to drive decisions regarding development and operation of the software concertedly with outside stakeholders, such as e.g., suppliers. Apart from consistent communication channels, in DevOps this requires thorough documentation of the software artefact, specially of operational characteristics and all runtime metadata \cite{ebert_scaling_2017}.
	\item \textbf{Cross-team integration:} Establishing inter-company collaboration \linebreak among teams of different domains to improve the software value stream \cite{forsgren_devops_2018} for the company. Apart from cross-discipline training of the software engineers, an unambiguous labeling of the software artefacts, a RACI (responsibility-assignment) matrix, and standardized documentation guidelines facilitate collaboration among teams \cite{erich_qualitative_2017}. Cross-functional teams shall not exceed the size of 8-12 so that they can maintain a focus on a specific aspect of the software \cite{dornenburg_path_2018}.
	\item \textbf{Skills growth:} Providing training materials and mentoring to acquire skills required for accomplishing the DevOps-related tasks. Having a supportive and visible support infrastructure on the level of concrete features and recurrent tools is critical for the wholesale success of software product \cite{dikert_challenges_2016}. Particularly for DevOps this requires establishing links between software artefacts, issue trackers, and deployment tools for capturing and documenting all relevant data throughout the lifecycle.
	\item \textbf{Cross-domain convergence:} Designing, developing, operating, and continuously improving software, so that it can be reused across application domains, e.g., mobile and pervasive sensing systems that share and process data among autonomous systems \cite{ebert_looking_2015}. Primarily this affects the selection of suitable network protocols for integrating the applications in the DevOps environment. Respective environments for this purpose must be able to deal with unreliable data connections and time-delayed delivery of operational data of the developed applications \cite{ebert_scaling_2017}.
\end{itemize}

\subsection{Themes and objectives behind NFRs of the business focus area}
\label{sec:business_themes}
Value capture, creation, and optimization throughout the software lifecycle is the prime objective behind business-focused NFRs in DevOps.
\begin{longtable}{P{190pt}P{62pt}P{185pt}}
\toprule
\textbf{Theme} & \textbf{Frequency} & \textbf{Primary studies} \\ \hline
Mean time to value & \centering 3 & \cite{dornenburg_path_2018,senapathi_devops_2018,somerville_fitting_2008} \\ \hline
Transaction costs & \centering 9 & \cite{adner_wide_2013,ameller_dealing_2019,aspara_creating_2013,erich_qualitative_2017,laanti_characteristics_2014,luoma_current_2012,parker_platform_2017,priem_insights_2011,riegel_systematic_2015} \\ \hline
Ecosystem valuation & \centering 9 & \cite{almeida_assessing_2015,bosch_integration_2010,bosch_softwares_2010,ebert_scaling_2017,fagerholm_right_2017,fricker_software_2012,hyrynsalmi_revenue_2012,olsson_holmstrom_ad_2017,rodriguez_continuous_2017} \\ \hline
Product market share & \centering 10 & \cite{boehm_roi_2004,bosch_speed_2016,forsgren_devops_2018,fricker_software_2012,huang_technology_2013,leber_value_2014,luo_qfd-based_2015,madachy_integrated_2006,raffo_integrating_2010,riegel_systematic_2015} \\ \hline
Value stream mapping & \centering 8 & \cite{bosch-mauchand_vcs:_2012,brenner_scaled_2015,ebert_looking_2015,forsgren_devops_2018,kersten_what_2018,kersten_mining_2018,paasivaara_adopting_2017,raffo_integrating_2010} \\ 
\bottomrule
\caption{Themes and objectives of business-focused NFRs.}
\label{tab:business_themes}
\end{longtable}

\begin{itemize}
	\item \textbf{Mean time to value:} Reducing the cycle time, i.e., the time between an idea and its realization as software in production. Operationally, the respective measures can be acquired from issue trackers and deployment systems \cite{dornenburg_path_2018}.
	\item \textbf{Transaction costs:} Balancing the software development and operation costs for the manufacturer with the perceived value of using the software for the customer. Also, customers conceive of costs in terms of the offer price plus all additional changes they need to undertake in order to use software \cite{adner_wide_2013}. The cost for the manufacturer accruing in the DevOps cycle can be calculated indirectly, e.g., through monitoring the frequency of bugs and code changes of particular features. Contrarily, the value of the software for the user can be evaluated using qualitative feedback methods (e.g., in-app surveys, ratings) that complement quantitative measures such as e.g., the invocation frequency or cause-effect analysis of customer behavior and offer price \cite{fagerholm_right_2017}.
	\item \textbf{Ecosystem valuation:} Assigning quantitative measures (either monetary or in form of a company balanced scorecard) to an ecosystem and its services \cite{almeida_assessing_2015}. A \textit{software ecosystem} consists of a software platform, internal/external developers and a community of domain experts. These domain experts help adapt or modify the software to satisfy the users’ needs \cite{bosch_integration_2010}. In DevOps, these quantitative measures needed for thorough ecosystem valuation can be acquired e.g., from ERP- and billing systems, bug trackers, development and operational systems. Overall, this allows to dynamically balance the company’s efforts for participation in the software ecosystem based on actual user behavior and thus optimize the business value of the software for the company.
	\item \textbf{Product market share:} Assessing and evaluating the impact of changes of the provisioned software on its market share, particularly through gap analysis between potential and actual market share, competitor analysis and value-estimation of new features \cite{forsgren_components_2018,madachy_integrated_2006}. Similar to ecosystem valuation, this theme requires quantification of external factors e.g., through mining textual product reviews, designing novel features, and weighting them according to estimated economic value added (EVA) and return on invest (ROI) \cite{boehm_roi_2004,luo_qfd-based_2015}. In DevOps, external factors such as e.g., the results from A/B testing, must then be considered in conjunction with internal factors such as e.g., business performance indicators, to decide upon concrete software engineering activities \cite{bosch_speed_2016}.
	\item \textbf{Value stream mapping:} Identifying and addressing delays and non-value-added activities in DevOps to accomplish the shortest sustainable lead time. The mapping between software artefacts and the business model’s value streams is the prerequisite for assessing the business value of software features, as also advocated in the Scaled Agile Framework (SAFe) \cite{paasivaara_adopting_2017}. For implementing this theme in DevOps, application lifecycle management (ALM) tools can be used to acquire the measures of the operational value stream, i.e., the steps used to provide goods or services to the customers. As an example, the invocation frequency of a feature in relation to the update cadence and its offer price can serve as an indicator for its value for the customer. These indicators can then be analyzed based on the cost and revenue model \cite{osterwalder_value_2014} to derive concrete activities for the development value stream, i.e., the steps used to create products, systems, or feature capabilities \cite{forsgren_devops_2018}.
\end{itemize}
\section{Discussion}
\label{sec:discussion}
In this section we summarize the principle findings from this systematic
mapping study and sketch their implications for researchers and
practitioners.

\subsection{Principle findings}

Our work outlines the broad scope and focus areas of NFRs, becoming
apparent through the pursed objectives originating from different
domains. Particularly, our study shows that NFRs to a large extent
address non-engineering qualities of software which are not covered
e.g., through the ISO/IEC \cite{international_organization_for_standardization_iso/iec_2018} which subsequently result in little if any
practical support for them in DevOps. This is problematic insofar, as
operational data accruing in the DevOps context, and generally in the
software development lifecycle, that are possibly suitable for guiding
software development is not utilized. Compared with the other focus
areas of NFRs, the majority of analyzed studies focus on technical
qualities, such as e.g., maintainability of the source code, safety, and
deployability. This may be due to the concrete definition of
technically-oriented NFRs in common literature and specifically in the
ISO/IEC specification \cite{international_organization_for_standardization_iso/iec_2018}. However, the importance of aligning software engineering,
organizational, and governance related objectives can also be inferred
from the number of studies elaborating on challenges arising from this
misbalance. As an example, the requirement for collecting operational
feature usage data is a prerequisite for A/B and feature testing \cite{olsson_requirements_2016,kevic_characterizing_2017} which must be implemented during
software development. However, the customer-focused NFR aiming on
\textit{feature-ideation from user interaction} (cf. Section \ref{sec:customer_themes}) itself
requires an overall product strategy and ecosystem valuation, so that
the acquired measures can be suitably evaluated. In addition, the
maintainability requirements for new software features derived from
actual user interaction might probably be higher than of features having
less strategic importance for a company. This example shows the required
indentation of different NFR focus areas, i.e., business, customer, and
development, to effectively build software in DevOps following the idea
of value-based software engineering \cite{boehm_value-based_2003}. In addition,
our study confirms the vagueness of the scope of "non-functional
requirements" itself, which has already been stressed by several authors
\cite{glinz_rethinking_2005,broy_rethinking_2015,kopczynska_empirical_2018}. Also, the alignment
between governance, regulatory compliance, and software development is
explicated as being crucial in many studies \cite{fricker_software_2012,adner_wide_2013,dikert_challenges_2016,yang_safety_2017} and the lack of methods, to address
these objectives in a balanced manner, underlined. Surprisingly many
studies elaborate on the human factor in software engineering and its
impact on the final quality of the software product, but neither present
applicable means to express, nor methods to measure and evaluate the
respective NFRs in DevOps.

\subsection{Implications for researchers and practitioners}

This study seeks to give an overview about typical focus areas, themes,
and objectives of NFRs in DevOps, to identify existing research gaps and
to sketch possible future research directions. Though NFRs being an
established concept in software engineering, there is a continuous and
rising interesting in this topic when it comes to the handling of NFRs
in DevOps, which also became apparent in the analysis of the publication
frequency (cf. Figure \ref{fig:year_trend}). While multiple studies describe (theoretical)
methods, models, and frameworks for handling NFRs in DevOps, few of them
are validated empirically. Also, besides classical engineering-oriented
NFRs, the automatic evaluation of NFRs arising from non-technical focus
areas in DevOps, e.g., focusing on business aspects, customers, and
governance requirements, is still hindered by the absence of a common
collection of metrics for these focus areas. These metrics also are the
prerequisite for balancing the objectives of the different NFRs focus
areas, e.g., to quantitatively express the allowed maintenance effort
for a software feature based on its usage. The operationalization of
non-engineering-focused NFRs and their evaluation, e.g., in relation to
a company's business model, plays a key role when it comes to
software-based digitalization in companies \cite{gruhn_bizdevops_2015,forbrig_integrating_2019,urbach_impact_2019}. As this study is based on
a literature analysis, the overall question remains which objectives are
actually important from a practical perspective and also how the results
from evaluating the NFRs of the diverse focus areas can be most
effectually utilized in DevOps.

In this frame, we are working on an approach for operationalizing and
evaluating NFRs in DevOps. Our approach is based on an operational
software quality model \cite{haindl_research_2019} to specify feature-dependent
NFRs, respective measures, and suitable instruments for their
acquisition in an automatable manner. The quality model is complemented
by a constraint definition language that utilizes these measures to
express complex NFRs and evaluate their fulfillment in DevOps.
Accordingly, the classification of focus areas, themes, and objectives
of NFRs presented in this study are the foundation for specifying NFRs
in DevOps through constraints and measures using this operational
quality model.

\section{Threats}
\label{sec:threats}
Though we conducted this systematic mapping study and the snowballing
procedure according to the respective guidelines \cite{petersen_systematic_2008,wohlin_guidelines_2014}, we recognize the following threats to the validity of the
study:

\begin{itemize}
\item \textbf{Publication bias:} Literature reviews are influenced by the common
    bias that positive research outcomes more likely are reported and
    published than negative ones. As the researched topics in this
    mapping study cannot be attributed to failure or success, we did not
    observe any noticeable bias among the reviewed studies in this
    regard. We contrarily recognized that multiple studies reflect on
    benefits of handling the multiform NFRs in DevOps, but also on
    illustrate the challenges and problems associated with these
    approaches. Overall, we recognized a well-balanced amount of studies
    reporting on either positive or negative aspects towards the
    examined topics. Hence, we consider the effect of this bias on our
    study as neglectable.
	\item \textbf{Lack of universal taxonomy:} While the term "non-functional
    requirement" is widely used in academic literature as well as in
    practice, in many cases a clear distinction between non-functional
    and functional requirements is difficult \cite{glinz_rethinking_2005,broy_rethinking_2015}.
    Also, many studies actually describe non-functional software
    requirements implicitly, particularly if these studies do not
    originate from the software engineering domain. To mitigate this
    threat, we conducted a snowballing procedure prior to the data
    analysis. Only if the primary studies cited in the initially
    retrieved papers were relevant for answering the research questions
    and adhered to the inclusion/exclusion criteria we included these
    studies in the subsequent analysis. Thus, we are confident that we
    captured the most relevant studies contributing to a broader
    understanding also of non-engineering focused NFRs in software
    projects.
	\item \textbf{Identification of primary studies:} We addressed this threat
    twofold: first, by composing the search string of multiple keywords
    each individually expressing a relevant aspect of NFRs in DevOps
    \cite{kitchenham_guidelines_2007} and second, by piloting this search
    string with different databases multiple times to evaluate the
    effect of the keywords onto the result set. However, the precision
    of the retrieved result set also depends on the quality of the
    keywords used in the searched studies. Also, we only used databases
    that cover the software engineering research very well and also read
    the titles, abstract, and conclusion (if needed) to decide upon the
    selection of a study. Thus, we are convinced to have captured the
    majority of the relevant studies and missed only little, if any,
    significant studies not matched by the search strings.
	\item \textbf{Study selection bias:} Though we are not aware of any biases
    regarding the selection of the studies, we recognize that unsuitable
    definition of the inclusion and exclusion criteria may lead to
    \textit{attrition bias}. To mitigate this threat, both researchers defined
    these criteria upfront in an unambiguous manner. Also, we revised
    these criteria during multiple when piloting the search string to
    assure to not exclude relevant studies. When deciding upon the
    selection of studies, we discussed any disagreements regarding the
    inclusion and exclusion of studies until both researchers agreed on
    a common decision.
	\item \textbf{Data extraction and analysis:} Biased judgments are also possible
    during data extraction and analysis, i.e., which text passages in
    the studies are labeled and which codes are derived and refined
    thereof. We mitigated this threat by defining the concrete
    properties to extract from the studies and the way to document this
    data upfront. Also, the thematic refinement of the codes and the
    encoded text fragments were discussed between both researchers until
    we agreed on a solid set of codes.
\end{itemize}
\section{Conclusion}
\label{sec:conclusion}
This paper presents the results from a systematic mapping study
examining the focus areas, themes, and objectives of NFRs in the DevOps
context. We retrieved 229 candidate studies from 5 academic databases,
of which 114 were selected as primary studies. Also, we included a
snowballing step to capture studies that elaborate on NFRs which not
adhere to the strict technically-oriented notion of this term but
nonetheless have an impact on the design, development, and operation of
a software system. This step again resulted in 27 primary studies being
selected for inclusion, with the result set totaling to 142 primary
studies for this mapping study. Data were extracted from these studies
and then categorized by a defined classification schema. Themes and
objectives of NFRs in DevOps described in the studies were iteratively
synthesized from labeled text fragments and then summarized per focus
area. The publication frequency over the years shows a continuous
interest in this field with higher increases since 2015, with the
studies being predominantly published in journals and on conferences.
Most studies in this field contribute theories, methods, models, or
present lessons learned from handling NFRs in DevOps in practical
settings. However, the majority of these studies describe artefacts
which are not yet practically implemented and thus lack an empirical
validation. From the studies, we condensed 7 recurrent focus areas of
NFRs in DevOps, ranging from development and operation to governance-
and business-oriented areas. The most pressing objectives pursued with
NFRs in DevOps are to assure safe and responsive operation of the
software (technical quality focus), facilitate cross-team integration of
experts from different domains (organization focus), ascertain
regulatory compliance of the software (governance focus), utilize
customer interaction with the software for prospective feature ideation
(customer focus), create and safeguard product market share (business
focus), optimize the delivery pipeline (development focus), and to
facilitate post-deployment debugging (operation focus).

For future work, typical metrics for the NFRs of the different focus
areas and usage scenarios in practice, as well as methods for their
operationalization shall be explored by researchers, at best
complemented with an empirical validation. Furthermore, methodological
and tool support is needed to aid practitioners specifying these NFRs in
a way that can be utilized in a DevOps context.

\pagebreak
\section*{Appendices}
\appendix
\section[Appendix A]{}
\label{appendix:appendix_a}
\centering
\begin{longtable}{P{110pt}p{340pt}}
\toprule
\textbf{Category} & \textbf{Description} \\ \hline
\multicolumn{2}{l}{\textit{P1: Research facet, adapted from Petersen \cite{petersen_systematic_2008} and Wieringa \cite{wieringa_requirements_2005}}} \\ \hline
Experience report & Similar to opinion paper, these works express practical experiences of the author with a method or technique. Often these reports come from practitioners who have used a specific method in practice and can thus provide valuable insights into possible ways for improvements. \\ \hline
Evaluation research & Describes the investigation of a problem or an implementation of a solution for it in practice, along with an evaluation for the solution. It also shows the benefits and drawbacks of the technique. \\ \hline
Validation research & The presented techniques are novel and have not yet been practically implemented, e.g., work done in laboratory. \\ \hline
Philosophical paper & The work sketches a new way on looking on things, such as taxonomies or concepts. Evaluation criteria are the originality, soundness, and transparency of the work. \\ \hline
Solution proposal & A solution technique for a problem is presented along with an argumentation for its relevance, but an evaluation of the solution is missing. The solution must be novel or demonstrate a significant enhancement of a previous approach. \\ \hline
Opinion paper & The work presents the personal opinion of the author about what is good or bad about a technique and how it can be improved. Such works do not rely on related works or sound research methodologies, but can possibly sketch new ways to address known problems. \\ \hline
\multicolumn{2}{l}{} \\ 
\multicolumn{2}{l}{\textit{P2: Contribution facet, adapted from Shaw \cite{shaw_writing_2003} and Paternoster \cite{paternoster_software_2014}}} \\ \hline
Model & An abstraction of an observed reality by concepts or related concepts after a conceptualization process. \\ \hline
Method & Method or technique for handling non-functional requirements in the planning, management, development, or operation of software. \\ \hline
Theory & Construct of cause-effect relationships of determined results. \\ \hline
Framework & Combination of methods with distinct conditions which method shall be applied under which circumstances and with defined input and output parameters. \\ \hline
Guideline & List of advises, compilation, or interpretation of obtained research results. \\ \hline
Lessons Learned & Commented set of outcomes with fact-based recommendations, directly concluded from the obtained research results. \\ \hline
Advice & Generic recommendation, influenced by personal opinions from the author. \\ \hline
Tool & Technology or software application used for handling non-functional requirements in the planning, management, development, or operation of software. \\
\bottomrule
\caption{Classification scheme for research and contribution facets.}
\label{tab:classification_scheme_facets}
\end{longtable}

\begin{longtable}{P{110pt}p{320pt}}
\toprule
\textbf{Category} & \textbf{Description} \\ \hline
\multicolumn{2}{l}{\textit{P5: Focus area of NFRs described in study}} \\ \hline
Customer & Studies describing NFRs related to increasing customer experience or value, e.g., through user-centered value estimation techniques or passive user involvement, e.g., through user monitoring.  \\ \hline
Development & Approaches capturing NFRs related to the software development, integration and delivery process and its tools, artefacts, methods, systems, and stakeholders. \\ \hline
Operation & Works that address NFRs affecting software operation and specifically includes the tools, artefacts, methods, systems, and involved stakeholders. \\ \hline
Governance & Presents liabilities, regulatory or technological compliances and requirement alignments that need to be taken account in DevOps and affect the NFRs of the respective software. \\ \hline
Quality & Works elaborating on or referring to static and dynamic quality factors of software that are linked with NFRs, e.g., the code or design quality, performance behavior or reliability. \\ \hline
Organization & Tackles NFRs attributed to the needed agility of the organization in the DevOps context to cope with internal or external changes. Particularly this comprises staff fluctuation, training, and (third party) supplier integration. \\ \hline
Business & Works regarding business-related NFRs that need to be taken into account in the DevOps context, e.g., to measure cost and revenue per user transaction or to evaluate fulfillment of business-driven goals with software. \\ 
\bottomrule
\caption{Classification schema for NFR focus areas.}
\label{tab:classification_scheme_focus_areas}
\end{longtable}
\section[Appendix B]{}
\label{appendix:appendix_b}
\centering
\begin{longtable}{P{40pt}p{90pt}p{90pt}p{96pt}p{100pt}}
\toprule
\textbf{\centering{Primary Study}} & \textbf{Research facet} & \textbf{Contribution facet} & \textbf{Focus area} & \textbf{Theme}\\ \hline
 
  \cite{forsgren_components_2018} & Experience Report&  Lessons Learned &  Customer &   Feature Ideation \\ \hline
  \cite{olsson_towards_2015} & Solution Proposal&  Framework &  Customer &   Feature Ideation \\ \hline
  \cite{dzamashvili_impact_2010} & Validation Research&Model  &  Customer &   Feature Ideation \\ \hline
  \cite{fabijan_customer_2015} &  Validation Research&Model  &  Customer &   Feature Ideation \\ \hline
  \cite{vanhala_role_2014} &Validation Research&Method  & Customer &   Feature Ideation \\ \hline
  \cite{fagerholm_building_2014} & Evaluation Research&Model  &  Customer &   Feature Ideation \\ \hline
  \cite{fagerholm_right_2017} & Validation Research&Model  &  Customer &   Feature Ideation \\ \hline
  \cite{erich_qualitative_2017} & Experience Report&  Lessons Learned &  Customer &   Feature Ideation \\ \hline
  \cite{bjarnason_challenges_2014} & Validation Research&Lessons Learned &  Customer &   Feature Ideation \\ \hline
  \cite{bosch_introducing_2011} & Validation Research&Guideline &  Customer &   Feature Ideation \\ \hline
  \cite{martinez-fernandez_towards_2018} &   Solution Proposal&  Framework &  Customer &   Feature Ideation \\ \hline
  \cite{rodriguez_continuous_2017} & Evaluation Research&Theory  & Customer &   Feature Ideation \\ \hline
  \cite{sauvola_towards_2015} &   Validation Research&Lessons Learned &  Customer &   Feature Ideation \\ \hline
  \cite{yaman_customer_2016}  & Validation Research&Theory  & Customer &   Feature Ideation \\ \hline
  \cite{fabijan_early_2015} &  Evaluation Research&Framework &  Customer &   Feature Ideation \\ \hline
  \cite{cao_agile_2008} &   Validation Research&Lessons Learned &  Customer &   Requirements Validation \\ \hline
  \cite{dzamashvili_impact_2010} &   Validation Research&Model  &  Customer &   Requirements Validation \\ \hline
  \cite{olsson_requirements_2016} & Solution Proposal&  Model  &  Customer &   Requirements Validation \\ \hline
  \cite{carrillo_metrics_2012} &  Validation Research&Theory  & Customer &   Value Creation \\ \hline
  \cite{huang_technology_2013} &Evaluation Research&Guideline &  Customer &   Value Creation \\ \hline
  \cite{kevic_characterizing_2017}  & Evaluation Research&Lessons Learned &  Customer &   Value Creation \\ \hline
  \cite{mehmood_evaluating_2009} &   Validation Research&Model  &  Customer &   Value Creation \\ \hline
  \cite{parker_platform_2017} &Philosophical Paper&Theory  & Customer &   Value Creation \\ \hline
  \cite{drutsa_practical_2015} &Validation Research&Theory  & Customer &   Value Creation \\ \hline
  \cite{riegel_systematic_2015} & Validation Research&Theory  & Customer &   Value Creation \\ \hline
  \cite{rodriguez_evaluation_2016} & Validation Research&Model  &  Customer &   Value Creation \\ \hline
  \cite{putta_benefits_2018}  & Validation Research&Theory  & Customer &   Release Cadence \\ \hline
  \cite{riungu-kalliosaari_devops_2016} &Validation Research&Lessons Learned &  Customer &   Release Cadence \\ \hline
  \cite{rodriguez_continuous_2017} & Evaluation Research&Theory  & Customer &   Release Cadence \\ \hline
  \cite{senapathi_devops_2018} & Experience Report&  Lessons Learned &  Customer &   Release Cadence \\ \hline
  \cite{dornenburg_path_2018} &   Experience Report&  Advice  & Customer &   Customer-Centered Epics Prioritization \\ \hline
  \cite{dingsoyr_exploring_2018} &  Validation Research&Advice  & Customer &   Customer-Centered Epics Prioritization \\ \hline
  \cite{erich_qualitative_2017}  & Experience Report&  Lessons Learned &Customer &   Customer-Centered Epics Prioritization \\ \hline
  \cite{favaro_managing_2002}  &   Philosophical Paper&Theory  & Customer &   Customer-Centered Epics Prioritization \\ \hline
  \cite{martinez-fernandez_quality_2018} &   Evaluation Research&Model  &  Customer &   Customer-Centered Epics Prioritization \\ \hline
  \cite{riegel_systematic_2015} & Validation Research&Theory  & Customer &   Customer-Centered Epics Prioritization \\ \hline
  \cite{senapathi_devops_2018} & Experience Report&  Lessons Learned &Customer &   Customer-Centered Epics Prioritization \\ \hline
  \cite{vahaniitty_towards_2008} &Philosophical Paper&Framework &  Customer &   Customer-Centered Epics Prioritization \\ \hline
  \cite{ameller_dealing_2019} &   Validation Research&Theory  & Governance & Enterprise Architecture \newline Alignment \\ \hline
  \cite{fleischmann_coherent_2010} &   Philosophical Paper&Tool  &   Governance & Enterprise Architecture \newline Alignment \\ \hline
  \cite{martinez-fernandez_towards_2018}&   Solution Proposal&  Framework &  Governance & Enterprise Architecture \newline Alignment \\ \hline
  \cite{shah_novel_2016} &   Solution Proposal&  Method  & Governance & Enterprise Architecture \newline Alignment \\ \hline
  \cite{urbinati_role_2018} &  Validation Research&Advice  & Governance & Enterprise Architecture \newline Alignment \\ \hline
  \cite{allix_androzoo:_2016}  & Experience Report&  Lessons Learned &Organization&   Market Responsiveness \\ \hline
  \cite{ebert_looking_2015}   &Opinion Paper &  Advice  & Organization&   Cross-Domain \newline Convergence \\ \hline
  \cite{ebert_scaling_2017} & Opinion Paper &  Lessons Learned &Organization&   Cross-Domain \newline Convergence \\ \hline
  \cite{fabijan_customer_2015} &  Validation Research&Model  &  Organization&   Cross-Domain \newline Convergence \\ \hline
  \cite{olsson_holmstrom_ad_2017} &   Evaluation Research&Method  & Organization&   Cross-Domain \newline Convergence \\ \hline
  \cite{regnell_supporting_2008} &   Solution Proposal&  Framework &  Organization&   Cross-Domain \newline Convergence \\ \hline
  \cite{robinson_roadmap_2010}  & Opinion Paper &  Framework &  Organization&   Cross-Domain \newline Convergence \\ \hline
  \cite{rodriguez_continuous_2017} & Evaluation Research&Theory  & Organization&   Cross-Domain \newline Convergence \\ \hline
  \cite{tontini_integrating_2007}  &  Evaluation Research&Method  & Organization&   Cross-Domain \newline Convergence \\ \hline
  \cite{gonzalez_calleros_towards_2014} &  Opinion Paper &  Theory  & Quality &Usability \\ \hline
  \cite{navarre_icos:_2009} &   Solution Proposal&  Method  & Quality &Usability \\ \hline
  \cite{rich_building_2009} & Opinion Paper &  Theory  & Quality &Usability \\ \hline
  \cite{cao_agile_2008} &   Validation Research&Lessons Learned &Operation &  Artefact \newline Runtime Traceability \\ \hline
  \cite{cito_context-based_2017} &  Validation Research&Method  & Operation &  Artefact \newline Runtime Traceability \\ \hline
  \cite{kersten_mining_2018}  & Solution Proposal&  Guideline &  Operation &  Artefact \newline Runtime Traceability \\ \hline
  \cite{lian_mining_2017}  &  Evaluation Research&Method  & Operation &  Artefact \newline Runtime Traceability \\ \hline
  \cite{niu_requirements_2018}  &   Philosophical Paper&Theory  & Operation &  Artefact \newline Runtime Traceability \\ \hline
  \cite{riegel_systematic_2015} & Validation Research&Theory  & Operation &  Artefact \newline Runtime Traceability \\ \hline
  \cite{bjarnason_challenges_2014} & Validation Research&Lessons Learned &Operation &  Post-Deployment \newline Debugging \\ \hline
  \cite{erich_qualitative_2017}  & Experience Report&  Lessons Learned &Operation &  Post-Deployment \newline Debugging \\ \hline
  \cite{guzman_how_2017} &Philosophical Paper&Model  &  Operation &  Post-Deployment \newline Debugging \\ \hline
  \cite{haleem_impact_2015} &   Philosophical Paper&Theory  & Operation &  Post-Deployment \newline Debugging \\ \hline
  \cite{luoma_current_2012}  & Validation Research&Guideline &  Operation &  Post-Deployment \newline Debugging \\ \hline
  \cite{martinez-fernandez_towards_2018}&   Solution Proposal&  Framework &  Operation &  Post-Deployment \newline Debugging \\ \hline
  \cite{niu_requirements_2018}  &   Philosophical Paper&Theory  & Operation &  Post-Deployment \newline Debugging \\ \hline
  \cite{vanhala_role_2014} &Validation Research&Method  & Operation &  Post-Deployment \newline Debugging \\ \hline
  \cite{di_nitto_software_2016} &  Solution Proposal&  Framework &  Operation &  Monitoring \newline Process Maturity  \\ \hline
  \cite{dikert_challenges_2016} & Validation Research&Theory  & Operation &  Monitoring \newline Process Maturity  \\ \hline
  \cite{kersten_mining_2018}  & Solution Proposal&  Guideline &  Operation &  Monitoring \newline Process Maturity  \\ \hline
  \cite{rodriguez_continuous_2017} &Validation Research&Lessons Learned &Operation &  Monitoring \newline Process Maturity  \\ \hline
  \cite{senapathi_devops_2018} & Experience Report&  Lessons Learned &Operation &  Monitoring \newline Process Maturity  \\ \hline
  \cite{shahin_beyond_2017} &Evaluation Research&Guideline &  Operation &  Monitoring \newline Process Maturity  \\ \hline
  \cite{di_nitto_software_2016} &  Solution Proposal&  Framework &  Development &Development Pipeline \newline Maturity \\ \hline
  \cite{forsgren_devops_2018} & Opinion Paper &  Guideline &  Development &Development Pipeline \newline Maturity \\ \hline
  \cite{jabbari_what_2016} &   Evaluation Research&Theory  & Development &Development Pipeline \newline Maturity \\ \hline
  \cite{kersten_mining_2018}  & Solution Proposal&  Guideline &  Development &Development Pipeline \newline Maturity \\ \hline
  \cite{rodriguez_continuous_2017} &Validation Research&Lessons Learned &Development &Development Pipeline \newline Maturity \\ \hline
  \cite{rodriguez_continuous_2017} & Evaluation Research&Theory  & Development &Development Pipeline \newline Maturity \\ \hline
  \cite{shahin_beyond_2017}  &Evaluation Research&Guideline &  Development &Development Pipeline \newline Maturity \\ \hline
  \cite{vargas_enabling_2018} &Solution Proposal&  Framework &  Development &Development Pipeline \newline Maturity \\ \hline
  \cite{cao_agile_2008} &   Validation Research&Lessons Learned &Development &Continuous Integration \\ \hline
  \cite{ebert_scaling_2017} & Opinion Paper &  Lessons Learned &Development &Continuous Integration \\ \hline
  \cite{kevic_characterizing_2017}  & Evaluation Research&Lessons Learned &Development &Continuous Integration \\ \hline
  \cite{shahin_beyond_2017}  &Evaluation Research&Guideline &  Development &Continuous Integration \\ \hline
  \cite{forsgren_software_2018} & Experience Report&  Process  &Development &Continuous Integration \\ \hline
  \cite{di_nitto_software_2016} &  Solution Proposal&  Framework &  Development &Continuous Integration \\ \hline
  \cite{ebert_scaling_2017} & Opinion Paper &  Lessons Learned &Development &Continuous Integration \\ \hline
  \cite{shahin_beyond_2017}  &Evaluation Research&Guideline &  Development &Continuous \newline Delivery \\ \hline
  \cite{ebert_scaling_2017} & Opinion Paper &  Lessons Learned &Development &Development Tool \newline Versatility \\ \hline
  \cite{gall_software_2014}  &  Experience Report&  Lessons Learned &Development &Development Tool \newline Versatility \\ \hline
  \cite{jabbari_what_2016} &   Evaluation Research&Theory  & Development &Development Tool \newline Versatility \\ \hline
  \cite{laanti_characteristics_2014}  &   Opinion Paper &  Theory  & Development &Development Tool \newline Versatility \\ \hline
  \cite{cao_agile_2008} &   Validation Research&Lessons Learned &Development &Development \newline Task Complexity \\ \hline
  \cite{dzamashvili_impact_2010} &   Validation Research&Model  &  Development &Development \newline Task Complexity \\ \hline
  \cite{karlsson_requirements_2007} &  Validation Research&Lessons Learned &Development &Development \newline Task Complexity \\ \hline
  \cite{kersten_mining_2018}  & Solution Proposal&  Guideline &  Development &Development \newline Task Complexity \\ \hline
  \cite{kopczynska_empirical_2018} &Validation Research&Theory  & Development &Development \newline Task Complexity \\ \hline
  \cite{martinez-fernandez_towards_2018}&   Evaluation Research&Model  &  Development &Development \newline Task Complexity \\ \hline
  \cite{ahmad_modeling_2015}  & Validation Research&Method  & Technical Quality&  Safety \\ \hline
  \cite{ameller_dealing_2019} &   Validation Research&Theory  & Technical Quality&  Safety \\ \hline
  \cite{berander_requirements_2005} & Philosophical Paper&Theory  & Technical Quality&  Safety \\ \hline
  \cite{blaine_software_2008} & Opinion Paper &  Advice  & Technical Quality&  Safety \\ \hline
  \cite{broy_rethinking_2015} & Philosophical Paper&Advice  & Technical Quality&  Safety \\ \hline
  \cite{ebert_looking_2015}   &Opinion Paper &  Advice  & Technical Quality&  Safety \\ \hline
  \cite{eckhardt_are_2016} &  Philosophical Paper&Lessons Learned &Technical Quality&  Safety \\ \hline
  \cite{herrmann_exploring_2007} &  Evaluation Research&Lessons Learned &Technical Quality&  Safety \\ \hline
  \cite{leber_value_2014} & Evaluation Research&Method  & Technical Quality&  Safety \\ \hline
  \cite{liu_extension_2009} &  Philosophical Paper&Method  & Technical Quality&  Safety \\ \hline
  \cite{ozkaya_making_2008} &Opinion Paper &  Advice  & Technical Quality&  Safety \\ \hline
  \cite{robinson_monitoring_2002} & Validation Research&Framework &  Technical Quality&  Safety \\ \hline
  \cite{rolland_modeling_2005} & Philosophical Paper&Theory  & Technical Quality&  Safety \\ \hline
  \cite{fagerholm_right_2017} & Validation Research&Model  &  Technical Quality&  Safety \\ \hline
  \cite{vickers_satisfying_2007} &  Opinion Paper &  Advice  & Technical Quality&  Safety \\ \hline
  \cite{vierhauser_developing_2015} &Evaluation Research&Framework &  Technical Quality&  Safety \\ \hline
  \cite{vierhauser_requirements_2016} &Validation Research&Theory  & Technical Quality&  Safety \\ \hline
  \cite{wagner_operationalised_2015} &Evaluation Research&Model  &  Technical Quality&  Safety \\ \hline
  \cite{zowghi_requirements_2005} &Philosophical Paper&Theory  & Technical Quality&  Safety \\ \hline
  \cite{adner_wide_2013}  &Opinion Paper &  Lessons Learned &Governance & Regulatory Compliance \\ \hline
  \cite{bjarnason_challenges_2014} & Validation Research&Lessons Learned &Governance & Regulatory Compliance \\ \hline
  \cite{boudreau_how_2009} & Opinion Paper &  Guideline &  Governance & Regulatory Compliance \\ \hline
  \cite{timmers_business_1998} &  Opinion Paper &  Theory  & Governance & Regulatory Compliance \\ \hline
  \cite{fahssi_enhanced_2015} &Solution Proposal&  Method  & Governance & Regulatory Compliance \\ \hline
  \cite{fricker_software_2012} &  Opinion Paper &  Theory  & Governance & Regulatory Compliance \\ \hline
  \cite{glinz_non-functional_2007}   &Philosophical Paper&Theory  & Governance & Regulatory Compliance \\ \hline
  \cite{kersten_what_2018} & Philosophical Paper&Theory  & Governance & Regulatory Compliance \\ \hline
  \cite{lian_mining_2017}  &  Evaluation Research&Method  & Governance & Regulatory Compliance \\ \hline
  \cite{parker_platform_2017}  &Philosophical Paper&Theory  & Governance & Regulatory Compliance \\ \hline
  \cite{zhang_data-driven_2017}  & Evaluation Research&Model  &  Governance & Regulatory Compliance \\ \hline
  \cite{zowghi_requirements_2005} &Philosophical Paper&Theory  & Governance & Regulatory Compliance \\ \hline
  \cite{hiisila_combining_2015} &   Experience Report&  Lessons Learned &Governance & Enterprise Architecture \newline Alignment \\ \hline
  \cite{svee_consumer_2012}  &  Evaluation Research&Theory  & Governance & Enterprise Architecture \newline Alignment \\ \hline
  \cite{bosch-sijtsema_user_2015} & Philosophical Paper&Framework &  Governance & Data Privacy \\ \hline
  \cite{di_nitto_software_2016} &  Solution Proposal&  Framework &  Governance & Data Privacy \\ \hline
  \cite{franch_data-driven_2018} &Evaluation Research&Lessons Learned &Governance & Data Privacy \\ \hline
  \cite{gall_software_2014}  &  Experience Report&  Lessons Learned &Governance & Data Privacy \\ \hline
  \cite{ivanov_implementation_2018} & Philosophical Paper&Framework &  Governance & Data Privacy \\ \hline
  \cite{kersten_what_2018}  & Philosophical Paper&Theory  & Governance & Data Privacy \\ \hline
  \cite{robinson_roadmap_2010}  & Opinion Paper &  Framework &  Governance & Data Privacy \\ \hline
  \cite{rodriguez_continuous_2017} & Evaluation Research&Theory  & Governance & Data Privacy \\ \hline
  \cite{shah_novel_2016} &   Solution Proposal&  Method  & Governance & Data Privacy \\ \hline
  \cite{vierhauser_requirements_2016} &Validation Research&Theory  & Governance & Data Privacy \\ \hline
  \cite{rodriguez_evaluation_2016} & Validation Research&Model  &  Technical Quality&  Usability \\ \hline
  \cite{griffith_industry_2017} &  Validation Research&Theory  & Technical Quality&  Resilience \\ \hline
  \cite{lamsweerde_handling_2000} &Evaluation Research&Method  & Technical Quality&  Resilience \\ \hline
  \cite{yang_actionable_2018}  &  Opinion Paper &  Advice  & Technical Quality&  Resilience \\ \hline
  \cite{ameller_dealing_2019} &   Validation Research&Theory  & Technical Quality&  Scalability \\ \hline
  \cite{fagerholm_building_2014} & Evaluation Research&Model  &  Technical Quality&  Scalability \\ \hline
  \cite{dingsoyr_exploring_2018} &  Validation Research&Advice  & Technical Quality&  Scalability \\ \hline
  \cite{putta_benefits_2018}  & Validation Research&Theory  & Technical Quality&  Scalability \\ \hline
  \cite{saadatmand_toward_2012} &Solution Proposal&  Method  & Technical Quality&  Scalability \\ \hline
  \cite{bosch_introducing_2011} & Validation Research&Guideline &  Technical Quality&  Performance \\ \hline
  \cite{chung_non-functional_2009} &  Validation Research&Lessons Learned &Technical Quality&  Performance \\ \hline
  \cite{cleland-huang_goal-centric_2005} & Validation Research&Method  & Technical Quality&  Performance \\ \hline
  \cite{fotrousi_quality_2014} &  Evaluation Research&Method  & Technical Quality&  Performance \\ \hline
  \cite{glinz_risk-based_2008} & Philosophical Paper&Method  & Technical Quality&  Performance \\ \hline
  \cite{klas_cqml_2009}  &  Validation Research&Model  &  Technical Quality&  Performance \\ \hline
  \cite{liu_extension_2009} &   Philosophical Paper&Method  & Technical Quality&  Performance \\ \hline
  \cite{matsumotoa_method_2017} &Philosophical Paper&Method  & Technical Quality&  Performance \\ \hline
  \cite{robinson_requirements_2006}  & Solution Proposal&  Framework &  Technical Quality&  Performance \\ \hline
  \cite{schmeling_composing_2011} & Solution Proposal&  Method  & Technical Quality&  Performance \\ \hline
  \cite{hecht_tracking_2015} & Validation Research&Tool  &   Technical Quality&  Framework Compliance \\ \hline
  \cite{kaiya_quality_2011} &Experience Report&  Method  & Technical Quality&  Framework Compliance \\ \hline
  \cite{r._plosch_collecting_2010} &Evaluation Research&Method  & Technical Quality&  Framework Compliance \\ \hline
  \cite{dikert_challenges_2016}  &Validation Research&Theory  & Governance & Regulatory Compliance \\ \hline
  \cite{ebert_scaling_2017} & Opinion Paper &  Lessons Learned &Governance & Regulatory Compliance \\ \hline
  \cite{fricker_software_2012} &  Opinion Paper &  Theory  & Governance & Regulatory Compliance \\ \hline
  \cite{glinz_non-functional_2007}   &Philosophical Paper&Theory  & Governance & Regulatory Compliance \\ \hline
  \cite{kevic_characterizing_2017}  & Evaluation Research&Lessons Learned &Governance & Regulatory Compliance \\ \hline
  \cite{kittlaus_software_2012}  & Opinion Paper &  Theory  & Governance & Regulatory Compliance \\ \hline
  \cite{parker_platform_2017}  &Philosophical Paper&Theory  & Governance & Regulatory Compliance \\ \hline
  \cite{riegel_systematic_2015} & Validation Research&Theory  & Governance & Regulatory Compliance \\ \hline
  \cite{rodriguez_continuous_2017} & Evaluation Research&Theory  & Governance & Regulatory Compliance \\ \hline
  \cite{schmeling_composing_2011} & Solution Proposal&  Method  & Governance & Regulatory Compliance \\ \hline
  \cite{zowghi_requirements_2005} &Philosophical Paper&Theory  & Governance & Regulatory Compliance \\ \hline
  \cite{carrillo_metrics_2012} &  Validation Research&Theory  & Technical Quality&  Security \\ \hline
  \cite{eckhardt_how_2015} &  Philosophical Paper&Method  & Technical Quality&  Security \\ \hline
  \cite{fabijan_customer_2015} &  Validation Research&Model  &  Technical Quality&  Security \\ \hline
  \cite{schief_business_2014} &   Philosophical Paper&Theory  & Technical Quality&  Security \\ \hline
  \cite{mehmood_evaluating_2009} &   Validation Research&Model  &  Technical Quality&  Extensibility \\ \hline
  \cite{yang_linking_2012} &  Philosophical Paper&Method  & Technical Quality&  Extensibility \\ \hline
  \cite{begel_analyze_2014} & Experience Report&  Advice  & Technical Quality&  Test Quality \\ \hline
  \cite{yang_safety_2017} &   Validation Research&Theory  & Technical Quality&  Test Quality \\ \hline
  \cite{olsson_holmstrom_ad_2017}&   Evaluation Research&Method  & Technical Quality&  Maintainability \\ \hline
  \cite{bosch_speed_2016}   &Opinion Paper &  Advice  & Technical Quality&  Maintainability \\ \hline
  \cite{lochmann_are_2013} &  Validation Research&Lessons Learned &Technical Quality&  Maintainability \\ \hline
  \cite{siavvas_qatch_2017} &   Evaluation Research&Framework &  Technical Quality&  Maintainability \\ \hline
  \cite{sauvola_towards_2015} &   Validation Research&Lessons Learned &Technical Quality&  Data-Driven Deployment \\ \hline
  \cite{olsson_requirements_2016} & Solution Proposal&  Model  &  Technical Quality&  Data-Driven Deployment \\ \hline
  \cite{olsson_towards_2015} & Solution Proposal&  Framework &  Technical Quality&  Data-Driven Deployment \\ \hline
  \cite{forsgren_measuring_2018} & Experience Report&  Process  &Organization&   Market Responsiveness \\ \hline
  \cite{bosch_introducing_2011} & Validation Research&Guideline &  Organization&   Market Responsiveness \\ \hline
  \cite{dikert_challenges_2016}  &Validation Research&Theory  & Organization&   Market Responsiveness \\ \hline
  \cite{bosch_integration_2010} &Philosophical Paper&Theory  & Organization&   Market Responsiveness \\ \hline
  \cite{laanti_characteristics_2014}  &   Opinion Paper &  Theory  & Organization&   Market Responsiveness \\ \hline
  \cite{putta_benefits_2018}  & Validation Research&Theory  & Organization&   Market Responsiveness \\ \hline
  \cite{raffo_integrating_2010} & Philosophical Paper&Guideline &  Organization&   Market Responsiveness \\ \hline
  \cite{yaman_customer_2016}  & Validation Research&Theory  & Organization&   Market Responsiveness \\ \hline
  \cite{foss_fifteen_2017} &   Validation Research&Theory  & Organization&   Resource Fluidity \\ \hline
  \cite{laanti_characteristics_2014}  &   Opinion Paper &  Theory  & Organization&   Resource Fluidity \\ \hline
  \cite{putta_benefits_2018}  & Validation Research&Theory  & Organization&   Resource Fluidity \\ \hline
  \cite{boudreau_how_2009} & Opinion Paper &  Guideline &  Organization&   Motivation \\ \hline
  \cite{bosch-sijtsema_user_2015} & Philosophical Paper&Framework &  Organization&   Motivation \\ \hline
  \cite{west_leveraging_2014} &  Philosophical Paper&Theory  & Organization&   Motivation \\ \hline
  \cite{ebert_scaling_2017} & Opinion Paper &  Lessons Learned &Organization&   Team-Supplier \newline Communication \\ \hline
  \cite{ebert_software_2014} &Opinion Paper &  Theory  & Organization&   Team-Supplier \newline Communication \\ \hline
  \cite{kittlaus_software_2012}  & Opinion Paper &  Theory  & Organization&   Team-Supplier \newline Communication \\ \hline
  \cite{lindberg_design_2011} &  Opinion Paper &  Advice  & Organization&   Team-Supplier \newline Communication \\ \hline
  \cite{senapathi_devops_2018} & Experience Report&  Lessons Learned &Organization&   Team-Supplier \newline Communication \\ \hline
  \cite{boehm_value-based_2003} &  Validation Research&Lessons Learned &Organization&   Cross-Team Integration \\ \hline
  \cite{bosch_introducing_2011} & Validation Research&Guideline &  Organization&   Cross-Team Integration \\ \hline
  \cite{ebert_scaling_2017} & Opinion Paper &  Lessons Learned &Organization&   Cross-Team Integration \\ \hline
  \cite{erich_qualitative_2017}  & Experience Report&  Lessons Learned &Organization&   Cross-Team Integration \\ \hline
  \cite{dornenburg_path_2018}  &   Experience Report&  Advice  & Organization&   Cross-Team Integration \\ \hline
  \cite{ebert_scaling_2017} & Opinion Paper &  Lessons Learned &Organization&   Cross-Team Integration \\ \hline
  \cite{forsgren_devops_2018} & Opinion Paper &  Guideline &  Organization&   Cross-Team Integration \\ \hline
  \cite{bosch_speed_2016}  &Opinion Paper &  Advice  & Organization&   Cross-Team Integration \\ \hline
  \cite{putta_benefits_2018}  & Validation Research&Theory  & Organization&   Cross-Team Integration \\ \hline
  \cite{wolbling_design_2012} &  Opinion Paper &  Theory  & Organization&   Cross-Team Integration \\ \hline
  \cite{ozkaya_making_2008}  &Opinion Paper &  Advice  & Organization&   Skills Growth \\ \hline
  \cite{senapathi_devops_2018} & Experience Report&  Lessons Learned &Organization&   Skills Growth \\ \hline
  \cite{ebert_software_2014} &Validation Research&Lessons Learned &Organization&   Skills Growth \\ \hline
  \cite{berander_requirements_2005} & Philosophical Paper&Theory  & Organization&   Skills Growth \\ \hline
  \cite{dikert_challenges_2016}  &Validation Research&Theory  & Organization&   Skills Growth \\ \hline
  \cite{erich_qualitative_2017}  & Experience Report&  Lessons Learned &Organization&   Skills Growth \\ \hline
  \cite{rodriguez_continuous_2017} &Validation Research&Lessons Learned &Organization&   Skills Growth \\ \hline
  \cite{dornenburg_path_2018}  &   Experience Report&  Advice  & Business &   Mean Time To Value \\ \hline
  \cite{senapathi_devops_2018} & Experience Report&  Lessons Learned &Business &   Mean Time To Value \\ \hline
  \cite{somerville_fitting_2008}  &   Opinion Paper &  Advice  & Business &   Mean Time To Value \\ \hline
  \cite{adner_wide_2013}   &Opinion Paper &  Lessons Learned &Business &   Costs \\ \hline
  \cite{ameller_dealing_2010} &   Validation Research&Theory  & Business &   Costs \\ \hline
  \cite{luoma_current_2012}  & Validation Research&Guideline &  Business &   Costs \\ \hline
  \cite{riegel_systematic_2015} & Validation Research&Theory  & Business &   Costs \\ \hline
  \cite{aspara_creating_2013} &  Validation Research&Model  &  Business &   Costs \\ \hline
  \cite{erich_qualitative_2017}  & Experience Report&  Lessons Learned &Business &   Costs \\ \hline
  \cite{laanti_characteristics_2014}  &   Opinion Paper &  Theory  & Business &   Costs \\ \hline
  \cite{parker_platform_2017}  &Philosophical Paper&Theory  & Business &   Costs \\ \hline
  \cite{priem_insights_2011} & Philosophical Paper&Theory  & Business &   Costs \\ \hline
  \cite{almeida_assessing_2015} &   Evaluation Research&Tool  &   Business &   Ecosystem Valuation \\ \hline
  \cite{bosch_softwares_2010} &Philosophical Paper&Theory  & Business &   Ecosystem Valuation \\ \hline
  \cite{ebert_scaling_2017} & Opinion Paper &  Lessons Learned &Business &   Ecosystem Valuation \\ \hline
  \cite{ebert_software_2014} &Opinion Paper &  Theory  & Business &   Ecosystem Valuation \\ \hline
  \cite{olsson_holmstrom_ad_2017}&   Evaluation Research&Method  & Business &   Ecosystem Valuation \\ \hline
  \cite{hyrynsalmi_revenue_2012} &Evaluation Research&Framework &  Business &   Ecosystem Valuation \\ \hline
  \cite{bosch_integration_2010} &Philosophical Paper&Theory  & Business &   Ecosystem Valuation \\ \hline
  \cite{rodriguez_continuous_2017} & Evaluation Research&Theory  & Business &   Ecosystem Valuation \\ \hline
  \cite{fagerholm_right_2017} & Validation Research&Model  &  Business &   Ecosystem Valuation \\ \hline
  \cite{boehm_roi_2004}  & Opinion Paper &  Theory  & Business &   Product Market Share \\ \hline
  \cite{huang_technology_2013}   &Evaluation Research&Guideline &  Business &   Product Market Share \\ \hline
  \cite{forsgren_devops_2018} & Opinion Paper &  Guideline &  Business &   Product Market Share \\ \hline
  \cite{ebert_software_2014} &Opinion Paper &  Theory  & Business &   Product Market Share \\ \hline
  \cite{bosch_speed_2016} &Opinion Paper &  Advice  & Business &   Product Market Share \\ \hline
  \cite{leber_value_2014} & Evaluation Research&Method  & Business &   Product Market Share \\ \hline
  \cite{madachy_integrated_2006} &  Opinion Paper &  Lessons Learned &Business &   Product Market Share \\ \hline
  \cite{raffo_integrating_2010}  & Philosophical Paper&Guideline &  Business &   Product Market Share \\ \hline
  \cite{riegel_systematic_2015} & Validation Research&Theory  & Business &   Product Market Share \\ \hline
  \cite{luo_qfd-based_2015}  &   Evaluation Research&Method  & Business &   Product Market Share \\ \hline
  \cite{bosch-mauchand_vcs:_2012} &Evaluation Research&Tool  &   Business &   Value Stream Mapping \\ \hline
  \cite{brenner_scaled_2015} &   Opinion Paper &  Advice  & Business &   Value Stream Mapping \\ \hline
  \cite{ebert_looking_2015}   &Opinion Paper &  Advice  & Business &   Value Stream Mapping \\ \hline
  \cite{forsgren_devops_2018} & Opinion Paper &  Guideline &  Business &   Value Stream Mapping \\ \hline
  \cite{kersten_mining_2018}  & Solution Proposal&  Guideline &  Business &   Value Stream Mapping \\ \hline
  \cite{kersten_what_2018}  & Philosophical Paper&Theory  & Business &   Value Stream Mapping \\ \hline
  \cite{paasivaara_adopting_2017}  &   Experience Report&  Lessons Learned &Business &   Value Stream Mapping \\ \hline
  \cite{raffo_integrating_2010}  & Philosophical Paper&Guideline &  Business &   Value Stream Mapping \\ 
\bottomrule
\caption{Systematic map overview.}
\label{tab:systematic map}
\end{longtable}

%\bibliographystyle{IEEEtran}
%\bibliography{manuscript}
\renewcommand{\refname}{Bibliography}
\printbibliography

\end{document}